\documentclass[final,5p,times]{elsarticle}
\usepackage{amsmath,amssymb,amsfonts}
\usepackage{algorithmic}
\usepackage{graphicx}
\DeclareGraphicsExtensions{.pdf,.jpeg,.png}
\usepackage{subfigure}
\usepackage{epsfig}
\usepackage{booktabs} %�������������������ݵļ���
\usepackage{multirow}
\usepackage{array}
\usepackage{dirtree}
\usepackage{epstopdf}
\usepackage{caption,setspace}
\usepackage{wrapfig}

\usepackage{graphics}

\journal{Journal of \LaTeX\ Templates}

\biboptions{sort&compress}%�ο����׿���ѹ����ʾ����1-3
\begin{document}

\begin{frontmatter}

\title{Reversible data hiding based on reducing invalid shifting of pixels in histogram shifting}
\tnotetext[mytitlenote]{This work was supported in part by the National Natural Science Foundation of China under Grant 61872003 , Grant U1636206, Grant 61672039 and Grant 61860206004.}

\author[1]{Yujie Jia}
\author[1]{Zhaoxia Yin\corref{mycorrespondingauthor}}
\ead{yinzhaoxia@ahu.edu.cn}
\cortext[mycorrespondingauthor]{Corresponding author}
\author[2]{Xinpeng Zhang}
\address[1]{Key Laboratory of Intelligent Computing $\&$ Signal Processing, Ministry of Education, Anhui University, Hefei 230601, P.R.China}
\address[2]{School of Computer Science, Fudan University, Shanghai 201203, P.R.China}
\author[3]{Yonglong Luo}
\address[3]{Anhui Provincial Key Laboratory of Network and Information Security, Wuhu, Anhui 240002, P.R.China}
\begin{abstract}
In recent years, reversible data hiding (RDH), a new research hotspot in the field of information security, has been paid more and more attention by researchers. Most of the existing RDH schemes do not fully take it into account that natural image's texture has influence on embedding distortion. The image distortion caused by embedding data in the image's smooth region is much smaller than that in the unsmooth region, essentially, it is because embedding additional data in the smooth region corresponds to fewer invalid shifting pixels (ISPs) in histogram shifting. Thus, we propose a RDH scheme based on the images
 texture to reduce invalid shifting of pixels in histogram shifting. Specifically, first, a cover image is divided into two sub-images by the checkerboard pattern, and then each sub-image's fluctuation values are calculated. Finally, additional data can be embedded into the region of sub-images with smaller fluctuation value preferentially. The experimental results demonstrate that the proposed method has higher capacity and better stego-image quality than some existing RDH schemes.
\end{abstract}

\begin{keyword}
%\texttt{elsarticle.cls}\sep \LaTeX\sep Elsevier \sep template
%\MSC[2010] 00-01\sep  99-00
Histogram shifting, invalid shifting pixels, reversible data hiding
\end{keyword}

\end{frontmatter}

%\linenumbers

\section{Introduction}
\label{sec::introduction}
\par As multimedia network technology develops rapidly, information security has attracted extensive attention in various fields. Data hiding refers to hiding data into the cover medium without being detected~\cite{zeng1998digital}. It is an effective technology to protect information. Data hiding methods mainly include steganography~\cite{lu2018secure,wang2018improving,wang2018joint,zhang2019binary}, digital watermarking~\cite{zhang2009fragile,qin2017fragile,yuan2018local,peng2018image}, construction based data hiding~\cite{otori2009texture,qian2017robust,li2019toward} and so on.
\par Most data hiding methods can accurately extract the embedded data but will cause some damage to the cover. However, for some specific scenarios, the distortion of the cover image is not tolerated, as is the slight distortion. In order to satisfy these specific occasions, a reversible data hiding (RDH) scheme was first proposed in 1997~\cite{barton1997method}. RDH means that not only the additional data can be extracted, but also the cover image can be reconstructed perfectly without any distortion. According to whether the cover image is encrypted, the RDH can be divided into two categories: RDH in encrypted domain and RDH in plaintext domain. Reversible data hiding in encrypted images (RDHEIs)~\cite{zhang2014reversibility,yi2017binary,yin2017reversible,yi2018parametric,yin2018reversible} means that the cover image is encrypted, while the RDH in plaintext domain does not encrypt the cover image and embeds the additional data in the plaintext domain directly.
\par In the past few decades, RDH schemes had been widely studied based on different techniques in the plaintext domain~\cite{shi2016reversible}, such as lossless compression~\cite{fridrich2001invertible,celik2005lossless,celik2006lossless,zhang2013recursive,qin2016reversible}, difference expansion (DE) ~\cite{tian2003reversible,hu2009based,li2013novel}, histogram shifting (HS)~\cite{ni2006reversible,tsai2009reversible,feng2012reversible,chen2013reversible,li2013general,lu2016asymmetric}, prediction error expansion (PEE)~\cite{thodi2007expansion,li2011efficient,li2013high,ou2013pairwise,qu2015pixel,jung2017high}, etc.
\begin{figure*}[!ht]
  \centering
  \subfigure[]{
   \label{fig-1-a}
    \includegraphics[width=0.3\textwidth]{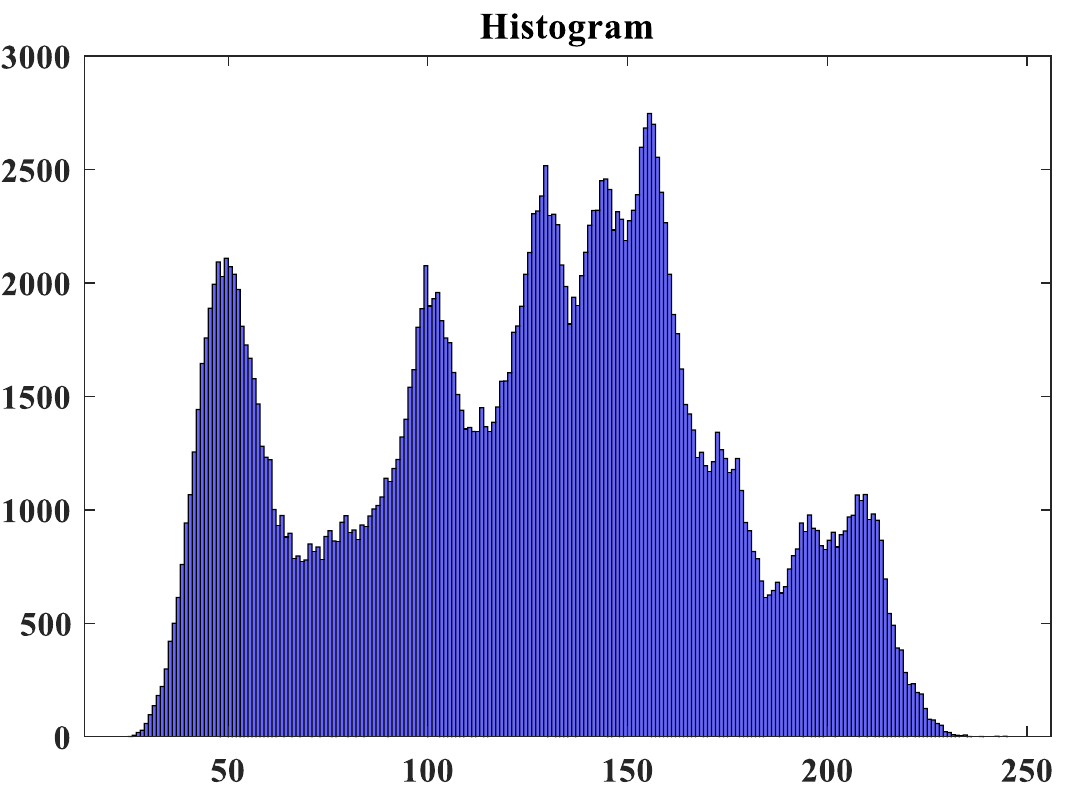}
  }
  \subfigure[]{
  \label{fig-1-b}
    \includegraphics[width=0.3\textwidth]{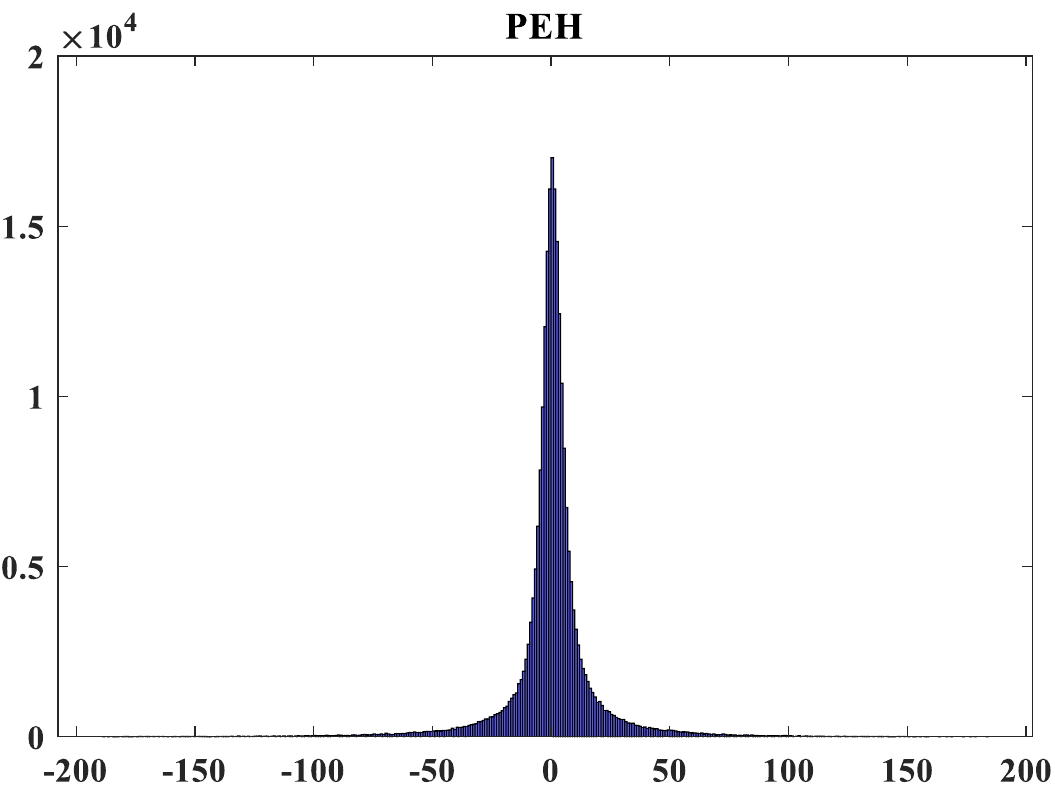}
  }
   \subfigure[]{
 \label{fig-1-c}
    \includegraphics[width=0.27\textwidth]{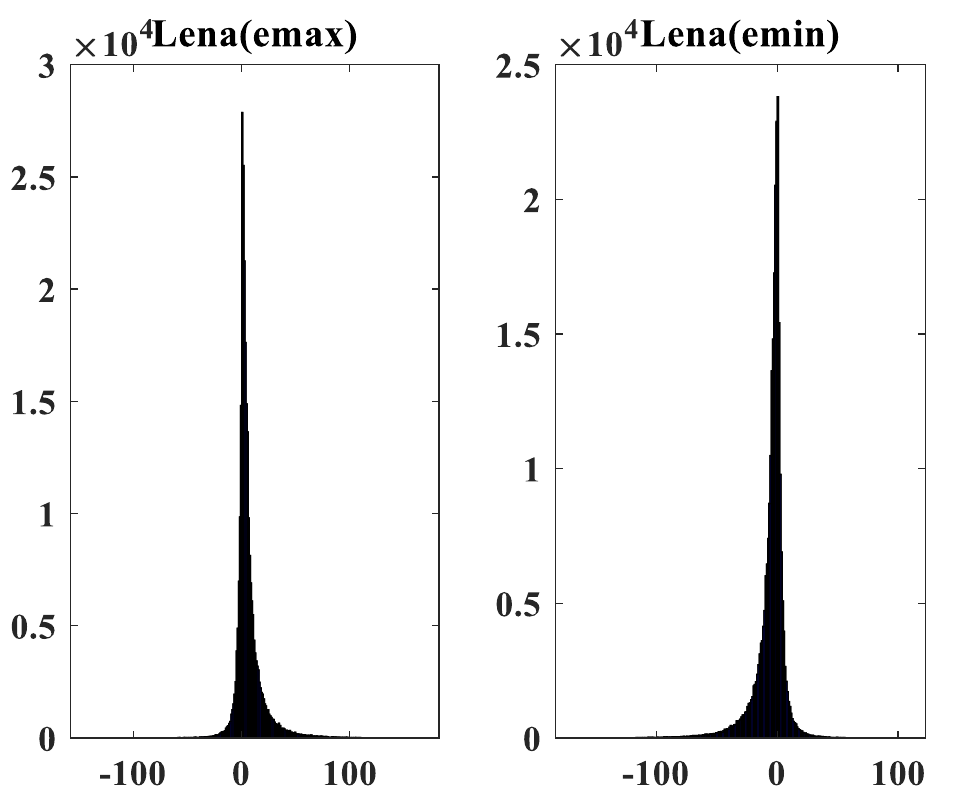}
 }
  \caption{The histograms generated by the methods~\cite{ni2006reversible},~\cite{tsai2009reversible},~\cite{chen2013reversible} on Lena. (a)Image histogram~\cite{ni2006reversible}. (b) PEH~\cite{tsai2009reversible}. (c)Asymmetric PEH~\cite{chen2013reversible}.}
 %\caption{The histograms generated by the methods [26], [27], [29] on Lena. (a)Image histogram [26]. (b) Prediction error histogram [27]. (c)Asymmetric PEH [29].}
\label{hs}
\end{figure*}
\begin{figure}[!ht]
\centering
  \includegraphics[width=0.4\textwidth]{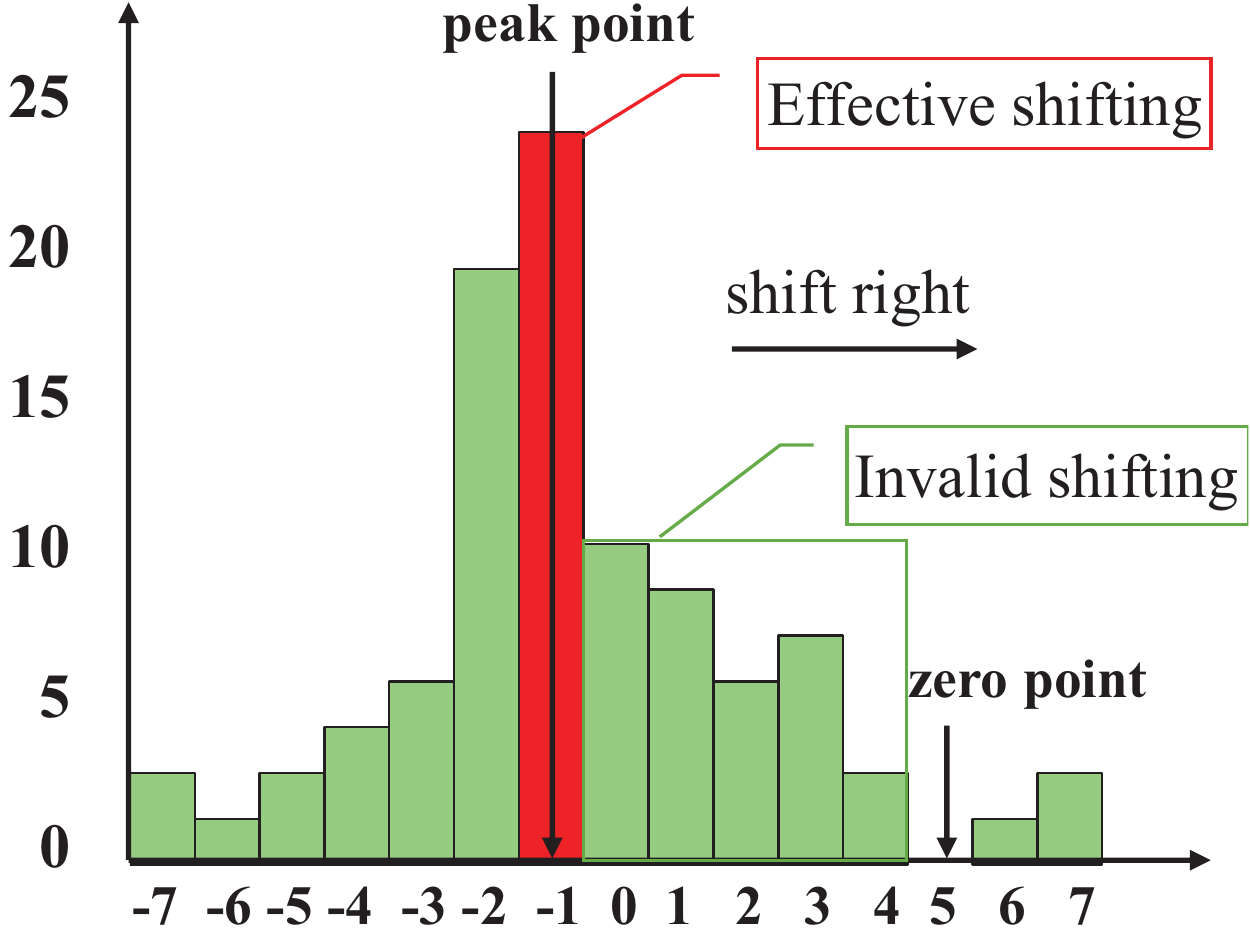}
 \caption{Example of pixels' valid shifting and invalid shifting.}
\label{invail}
\end{figure}
\begin{figure*}[!ht]
  \centering
    \includegraphics[width=0.8\textwidth]{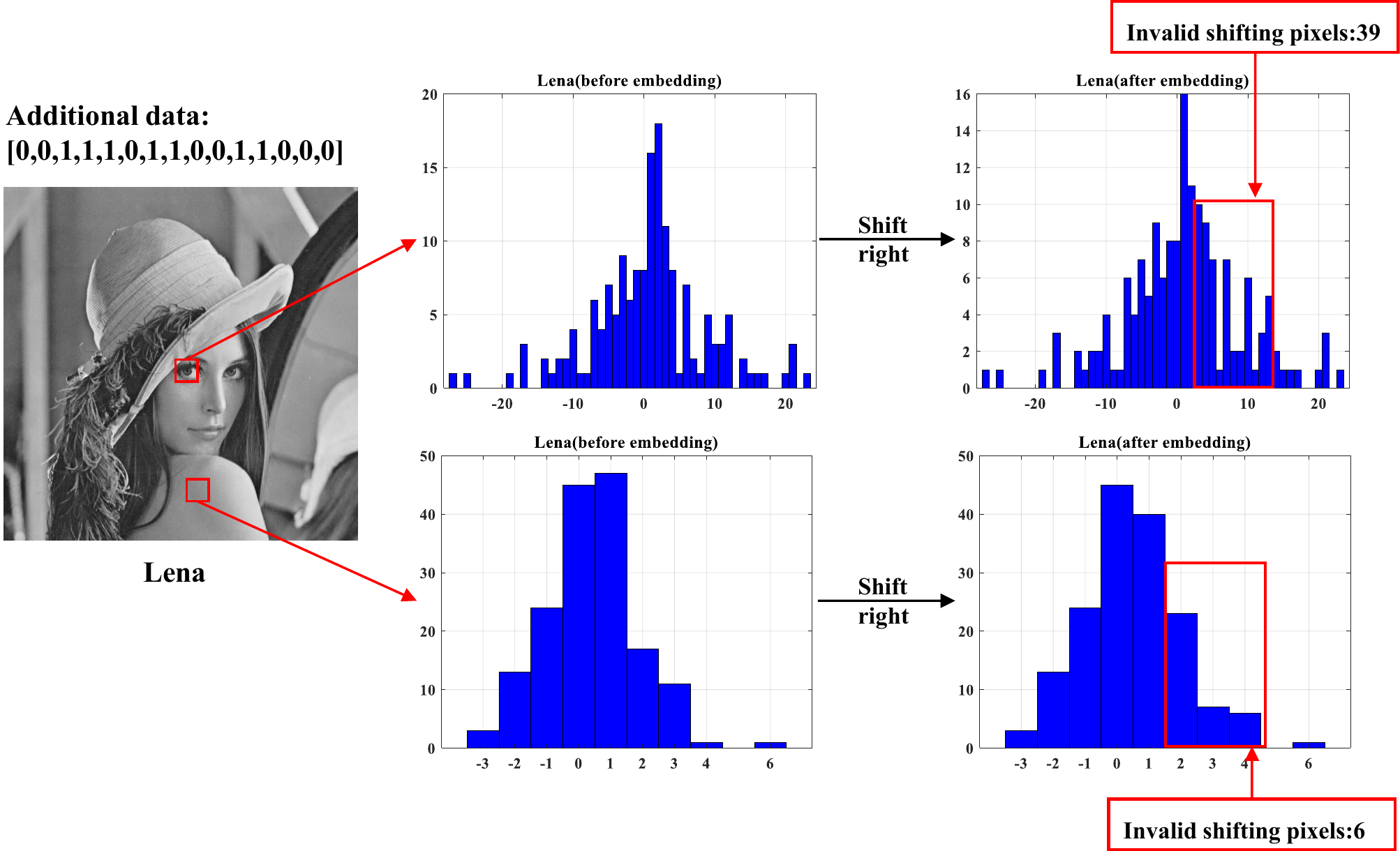}
  \caption{Comparison of the number of ISPs after embedding the same additional data in the PEHs of unsmooth region and smooth region with the same size in Lena.}
\label{smooth}
\end{figure*}
\par The lossless compression algorithm~\cite{fridrich2001invertible} is the most common method in early RDH. The main idea is to save space for hiding additional data by performing lossless compression on the cover image. In 2016, Qin et al.~\cite{qin2016reversible} improved the search order coding based on VQ and proposed an adaptive switching method. Their approach has greatly improved embedding capacity compared to some methods. Obviously, this algorithm relies on the performance of the lossless compression algorithm, and its embedding capacity will be limited due to the weak correlation between bit planes. In addition, direct compression of the cover image will produce noise, which will degrade the image quality. Therefore, the performance of the RDH scheme based on bit plane compression is not satisfactory. Shortly afterwards, DE ~\cite{tian2003reversible} was proposed to meet higher embedding capacity and better visual quality, and the basic idea of DE is to expand the difference between a pair of pixels to reversibly embed additional data.
\par In recent years, algorithms based on HS are most commonly used. HS~\cite{ni2006reversible} is a typical RDH method that embeds additional data into the peak point pixels of the image gray histogram and shifts others, and HS is put forward by Ni et al. in 2006. The main idea is to embed additional data by keeping the peak point pixels unchanged to hide '0' and shifting the peak point pixels by 1 to hide '1', and shift pixels which are between peak point and zero point by adding 1 or subtracting 1. As long as the peak point is known, the process of extraction and recovery can be easily accomplished. However, the embedding capacity of this method is relatively low. Later, in 2009, Tsai et al.~\cite{tsai2009reversible} proposed a method to construct prediction error histogram(PEH) including the negative PEH and non-negative PEH, which effectively increases the embedding capacity. Since the distribution of the prediction error value is more compact than that of the image pixel value, the PEH is steeper than the image pixels histogram and the capacity can be effectively improved. Afterwards, a method of constructing asymmetric histograms is proposed to embed additional data, which is characterized by the compensation of secondary embedding. Chen et al.~\cite{chen2013reversible} used the maximum prediction error and the minimum prediction error of the image to form an asymmetric PEH in 2013. Firstly, embed additional data at the peak point of the maximum PEH, and the pixel value between the peak point and the left zero point subtracts 1. Then, the additional data is embedded at the peak point of the minimum PEH, and the pixel value between the peak point and the right zero point is added by 1. The distortion of the image is reduced by the complementary effect of pixels generated in the embedding process. Fig.~\ref{hs} shows the histograms generated by the methods ~\cite{ni2006reversible},~\cite{tsai2009reversible},~\cite{chen2013reversible} on Lena, which are image histogram, prediction error histogram, and asymmetric histograms. It can be seen from Fig.~\ref{hs}(a)(b) that the prediction error histogram is steeper and the frequency of the peak point in the prediction error histogram is much higher than that in the image histogram. In addition, Fig.~\ref{hs}(c) constructs asymmetric histograms, which use secondary embedding at the same position to achieve compensation, effectively improving the embedding capacity and reducing image distortion.
\par PEE~\cite{thodi2007expansion}, as another commonly used and effective method, embeds the additional data into prediction error employing DE, and PEE could be seen as a special HS~\cite{li2013general}. Subsequently, combining with PEE, the method for achieving better embedding efficiency by pixel value ordering (PVO) was proposed by Li et al~\cite{li2013high}, which can achieve embedding two bits of additional data for every four pixels. In 2017, Jung K H~\cite{jung2017high} made improvements based on the method of Li et al.~\cite{li2013high}, including embedding capacity and image quality. Jung K H first divided the image into $3\times1$ non-overlapping blocks, then predicted the maximum and minimum values within the block, and finally embedded the secret data and recovered the image by the method of PEE. Jung K H's algorithm achieves embedding two bits of additional data for every three pixels.
\par It can be found that most current methods, such as PEH, PEE or PVO, mostly embed additional data in a certain scanning order, without considering the texture information of the image. Fundamentally, the effect of the number of invalid shifting pixels (ISPs) in the HS on image distortion is not considered. Here, the ISPs means the pixel shifted but not selected for embedding. Fig.~\ref{invail} gives an example of pixels' valid shifting and invalid shifting. For a series of additional data, the number of valid shifting pixels produced by histogram shifting is fixed during the embedding process, so that the distortion of the image is largely affected by the number of ISPs. That is, the more the number of ISPs, the greater the image distortion. Here, Fig.~\ref{smooth} shows the comparison of the numbers of ISPs in the unsmooth region and smooth region with the same size in Lena, and the same additional data is embedded in the PEHs of the two regions. The unsmooth region and the smooth region selected are the eye and shoulder, respectively, where the size of the block is $20\times20$, and the number of additional data is 15 bits. It has been experimentally found that the number of ISPs obtained in the unsmooth region is 39, and that in the smooth region is 6. In order to facilitate an intuitive comparison, we marked the number of ISPs in each area in the Fig.~\ref{smooth}. Based on this, in this paper, we propose a RDH scheme based on reducing the invalid shifting of pixels in the histogram shifting, which is achieved by preferentially embedding additional data in the smooth region of the image. Specifically, Firstly, we divide the pixel values in the image in a checkerboard manner. Secondly, we calculate the fluctuation value of each pixel and sort them. Finally, we combine the prediction error and the fluctuation to prioritize the embedding of additional data in small fluctuations. The experimental results of the proposed scheme show that not only the distortions effectively reduced but also the capacity is greatly improved.
\par The structure of the rest paper is as follows. The detailed process of the proposed scheme is introduced in Section~\ref{sec::Proposed method}. Next, Section~\ref{sec::Experimental} presents the results and analysis of the experiment. Finally, we summarize the paper in Section ~\ref{sec::Conclusion}.

\section{Proposed method}
\label{sec::Proposed method}
In this section, we will describe the proposed method in detail. Firstly, the detailed process of calculating the fluctuation value is introduced. Secondly, give the process of how to calculate the prediction error. Finally, the procedure of embedding and extracting is given.
\subsection{Calculation of fluctuation value}
\label{subsec::fluctuation values}
\par
\begin{figure}[!ht]
\centering
  \includegraphics[width=0.2\textwidth]{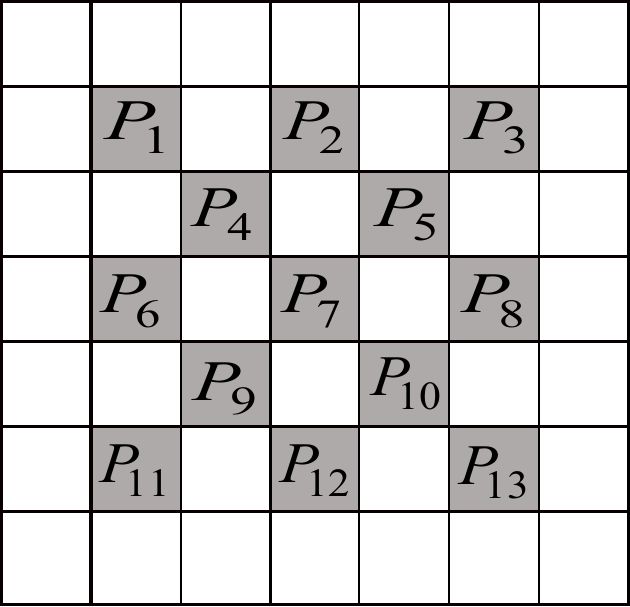}
 \caption{An example of allocating pixels with checkerboard, the gray and white belong to set A and set B respectively.}
\label{AB}
\end{figure}

 As shown in Fig.~\ref{AB}, the image is divided into two sets, A and B, in a checkerboard pattern, regardless of the pixels on the boundary, set A is gray while set B is white. Since the calculation methods of the fluctuation values in A and B are the same, here, we will take A as an example to introduce the specific scheme of calculating the fluctuation value.
\begin{figure}[!ht]
  \centering
    \includegraphics[width=0.1\textwidth]{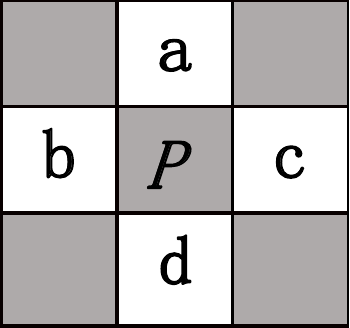}
  \caption{The adjacent pixels of a pixel.}
\label{AB1}
\end{figure}
\begin{figure*}[!ht]
  \centering
    \includegraphics[width=0.8\textwidth]{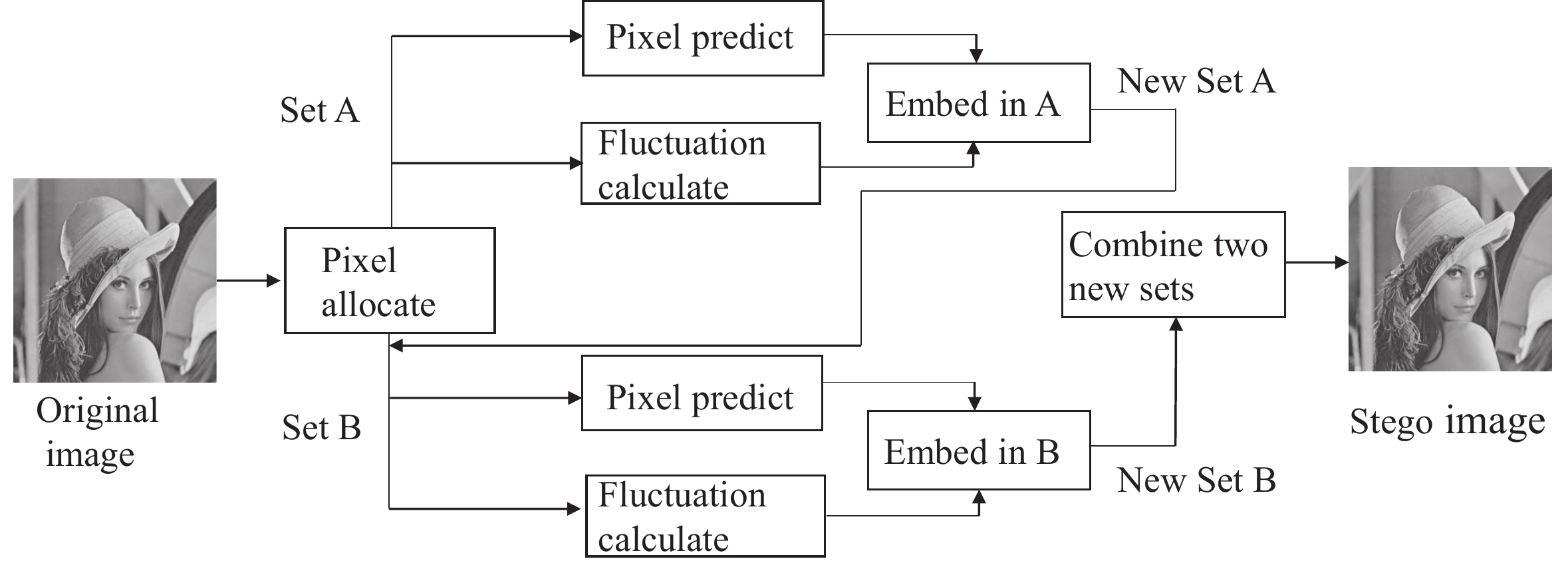}
  \caption{Framework of data embedding procedure.}
\label{framework}
\end{figure*}
\begin{enumerate}
\item Calculate the local complexity of each pixel in A, the formula is as follows:
\begin{equation}
\label{eq::eq1}
\centering
\ \Omega _{p}\!=\!\left | a\!-\!d \right |\!+\!\left | b\!-\!c \right |\!+\!\left | a\!+\!c\!-\!b\!-\!d \right |\!+\!\left | c\!+\!d\!-\!a\!-\!b \right | \text{,}
\end{equation}
\par where $a$, $b$, $c$, and $d$ are the upper, left, right, and lower pixel values adjacent to the pixel point $P$ as shown in Fig.~\ref{AB1}. The four terms in the formula correspond to the changes between four nearest pixel neighbors of $P$ along vertical, horizontal, positive diagonal and negative diagonal direction respectively.
\item In order to measure the smoothness of the pixel more accurately, we calculate the fluctuation value F of the current pixel value in combination with the local complexity of its adjacent pixel values. Taking the pixel value in Fig.~\ref{AB} as an example, the computing of the fluctuation value F of the pixel point $P_{1}$ is following:
\begin{equation}
\label{eq::eq2}
\centering
\ F_{P_{1}}=\Omega _{P_{1}}+\Omega _{P_{4}}\text{.}
\end{equation}
\par Calculate the fluctuation value F of the pixel point $P_{2}$ in Fig.~\ref{AB} as:
\begin{equation}
\label{eq::eq3}
\centering
\ F_{P_{2}}=\Omega _{P_{2}}+\left \lfloor \frac{\Omega _{P_{4}}+\Omega _{P_{5}}}{2}\right \rfloor\text{.}
\end{equation}
\par Similarly, the calculation of the fluctuation of the pixel $P_{4}$ in Fig.~\ref{AB} should be:
\begin{equation}
\label{eq::eq4}
\centering
\ F_{P_{4}}=\Omega _{P_{4}}+\left \lfloor  \frac{\Omega _{P_{1}}+\Omega _{P_{2}}+\Omega _{P_{6}}+\Omega _{P_{7}}}{4} \right   \rfloor\text{.}
\end{equation}
\par The above can be concluded that when there is one, two, or four adjacent pixels of the pixel in A, their fluctuation values are gained by the equations ~\eqref{eq::eq2}, ~\eqref{eq::eq3}\ , or~\eqref{eq::eq4}, respectively.
\end{enumerate}
\begin{figure*}[!ht]
  \centering
  \subfigure[]{
   \label{fig-7-a}
    \includegraphics[width=0.4\textwidth]{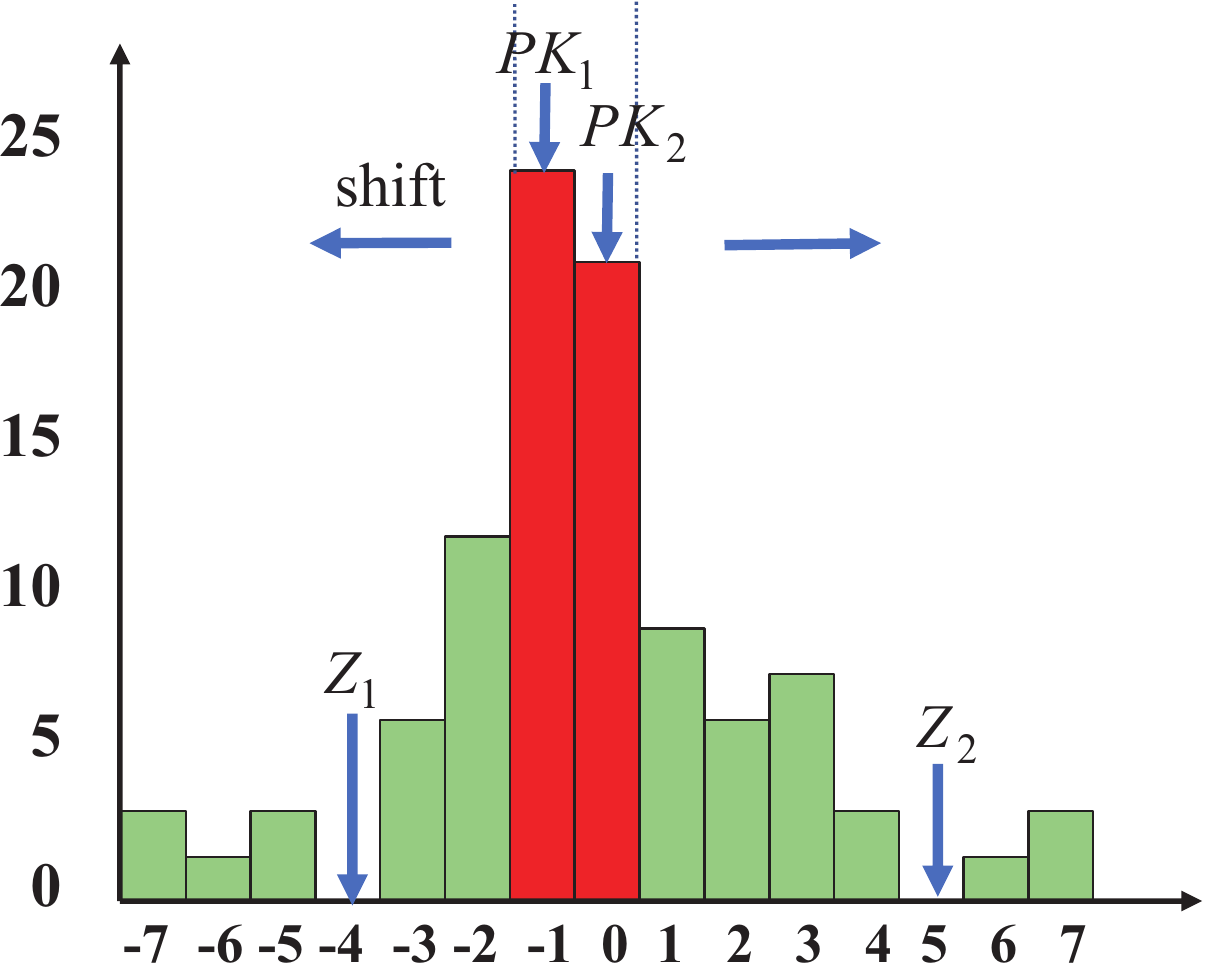}
  }
  \subfigure[]{
  \label{fig-7-b}
    \includegraphics[width=0.4\textwidth]{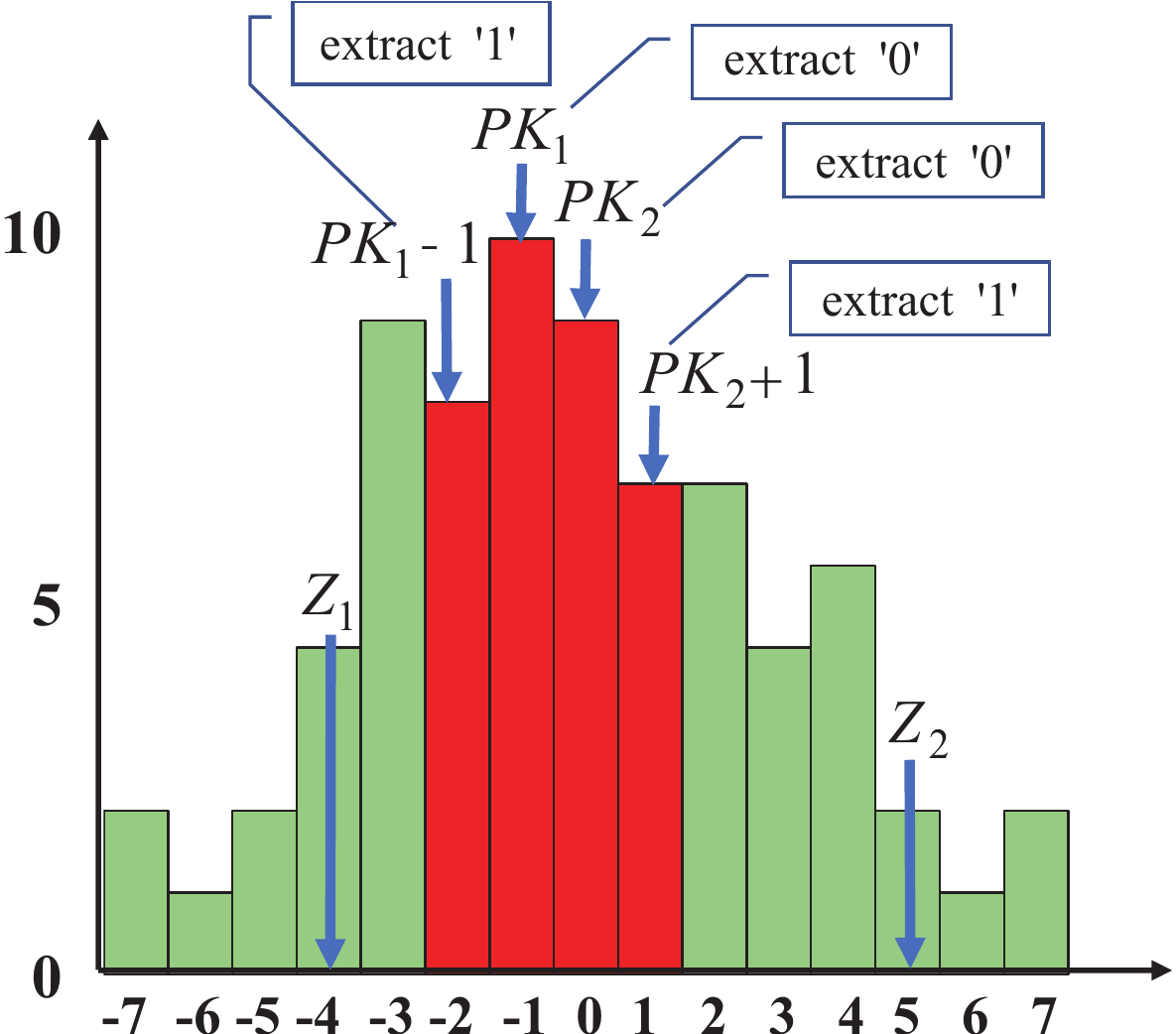}
  }
  \caption{Examples of histogram shifting in embedding and extraction. (a) Embedding. (b) Extraction.}
\label{emd-ext}
\end{figure*}

\subsection{Calculation of prediction error}
\label{subsec::prediction error}
\par Since the processes of A and B layers are similar, we still cite A as an instance to illustrate how to calculate the prediction error. For each pixel, we use four nearest neighbors to predict it.
\begin{equation}
\label{eq::eq5}
\centering
\ {P}''_{u,v}=\omega _{1}\!\cdot\! P_{u-1,v}\!+\!\omega _{2}\!\cdot\! P_{u+1,v}\!+\!\omega _{3}\!\cdot\! P_{u,v-1}\!+\!\omega _{4}\!\cdot\! P_{u,v+1},
\end{equation}
where ${P}''$ represents the predicted value of the pixel value $P_{u,v}$, $u$ and $v$ are the abscissa and ordinate of the pixel value $P_{u,v}$, respectively, and $\omega _{1}$, $\omega _{2}$, $\omega _{3}$, $\omega _{4}$ are the weights of the corresponding positions, where $\omega _{1}+\omega _{2}+\omega _{3}+\omega _{4}=1$.
\par The prediction error is calculated by
\begin{equation}
\label{eq::eq6}
\centering
\ e_{u,v}=P_{u,v}-{P}''_{u,v}\text{.}
\end{equation}
\par The solution to the four weights $\omega _{i}$ in ~\eqref{eq::eq5} is shown below:
\begin{enumerate}
\item Calculate the mean of the four nearest neighbors of each pixel value in A.
\begin{equation}
\label{eq::eq7}
\centering
\ {P}'_{u,v}\!=\!\left \lfloor \frac{P_{u-1,v}\!+\!P_{u+1,v}\!+\!P_{u,v-1}\!+\!P_{u,v+1}}{4} \right \rfloor\text{,}
\end{equation}
\par where $\left \lfloor \cdot  \right \rfloor$ represents the floor function.
\item The difference between ${P}'_{u,v}$ and the nearest neighbors is obtained by subtracting the values of the upper, lower, left and right, respectively, and expressed as $e_{1}$, $e_{2}$, $e_{3}$, and $e_{4}$. The corresponding $\omega _{i}$ is calculated as follows:
\begin{equation}
\label{eq::eq8}
\centering
\ {\omega }'_{i}=\begin{cases}
\frac{1}{4} & \text{ if } \sum_{i=1}^{4} \left | e_{i} \right |=0 \\
\frac{\sum _{i=1}^{4}\left | e_{i} \right |}{1+\left | e_{i} \right |}& \text{ otherwise }\text{,}
\end{cases}
\end{equation}
\par where $i\epsilon \left [ 1,4 \right ]$, normalize ${\omega }'_{1}$, ${\omega }'_{2}$, ${\omega }'_{3}$, and ${\omega }'_{4}$ to get $\omega _{1}$, $\omega _{2}$, $\omega _{3}$, and $\omega _{4}$ which are the four weights of upper, lower, left, and right, respectively.
\end{enumerate}

\subsection{Embedding procedure}
\label{subsec::Embedding}
\par Fig.~\ref{framework} shows the framework of data embedding procedure. In order to avoid the underflow/overflow problem during embedding, regardless of boundary pixels, we modify the pixel with value of 0 or 255 which may occur underflow/overflow during embedding or shifting, the specific operation is that change 0 to 1 and change 255 to 254, and mark 1 in the location map. In addition, for other pixels, if they are modified to 0 or 255 due to embedding or shifting, we mark 0 in the location map. Then, location map is lossless compressed to a smaller size as part of the payload. The detailed embedding process is depicted as:
\begin{enumerate}
\item The fluctuation values and prediction errors of all pixels in A are acquired according to B, and then the fluctuation sequence and the prediction error sequence are obtained in the same scanning order. Sort the fluctuation sequence in an ascending order to gain the sequence $\left\{\begin{matrix}
F_{A_{1}},F_{A_{2}},...,F_{A_{i}}
\end{matrix}\right\}$ and the sequence $\left \{ e_{F_{A_{1}}},e_{F_{A_{2}}},...,e_{F _{A_{i}}} \right \}$  is got by sorting the prediction error sequence in ascending order of the fluctuation values.
\item Allocate half of the payload to A. We modify the prediction error in sequence  $\left \{ e_{F_{A_{1}}},e_{F_{A_{2}}},...,e_{F _{A_{i}}} \right \}$  in turn to embed additional data. The regulation is:
\begin{small}
\begin{equation}
\label{eq::eq9}
\centering
\ e_{i}^{'}=\left\{\begin{array}{ll}
e_{i}+b & \text{ if } e_{i}=\text{max}\left ( PK_{1},PK_{2} \right ) \\
e_{i}-b & \text{ if } e_{i}=\text{min}\left ( PK_{1},PK_{2} \right ) \\
\multirow{2}*{$e_{i}+1$} & \text{ if } e_{i}\!>\!\text{max}\left ( PK_{1},PK_{2} \right )\ \text{,} \\& and\ e_{i}\!<\!\text{max}\left ( Z_{1},Z_{2} \right ) \\
\multirow{2}*{$e_{i}-1$} & \text{ if } e_{i}\!<\!\text{min}\left ( PK_{1},PK_{2} \right )\ \\& and\ e_{i}\!>\!\text{min}\left ( Z_{1},Z_{2} \right )\\
e_{i} & \text{ otherwise }
\end{array}\right.
\end{equation}
\end{small}
\par where $b$ is additional data, $b\epsilon \left \{ 0,1 \right \}$, $PK_{1}$ and $PK_{2}$ are peak points when embedding, in addition, $Z_{1}$ and $Z_{2}$ are two zero points closest to the peak points. $e_{i}^{'}$ is the marked prediction error, and the corresponding original pixel value is modified by:
\begin{equation}
\label{eq::eq10}
\centering
\ P_{i}^{'''}=P_{i}^{''}+e_{i}^{'}\text{,}
\end{equation}
\par where $P_{i}^{''}$ is the predicted value, $e_{i}^{'}$ is the corresponding marked prediction error, and $P_{i}^{'''}$ is the marked pixel value.
\item According to ~\eqref{eq::eq10}, the marked image of A is obtained, and on the basis of it, we implement the embedding of B in the same manner.
\end{enumerate}

\subsection{Extraction and recovery procedure}
\label{subsec::Extraction}
\begin{figure*}[!ht]
  \centering
  \subfigure[]{
   \label{fig-8-a}
    \includegraphics[width=0.2\textwidth]{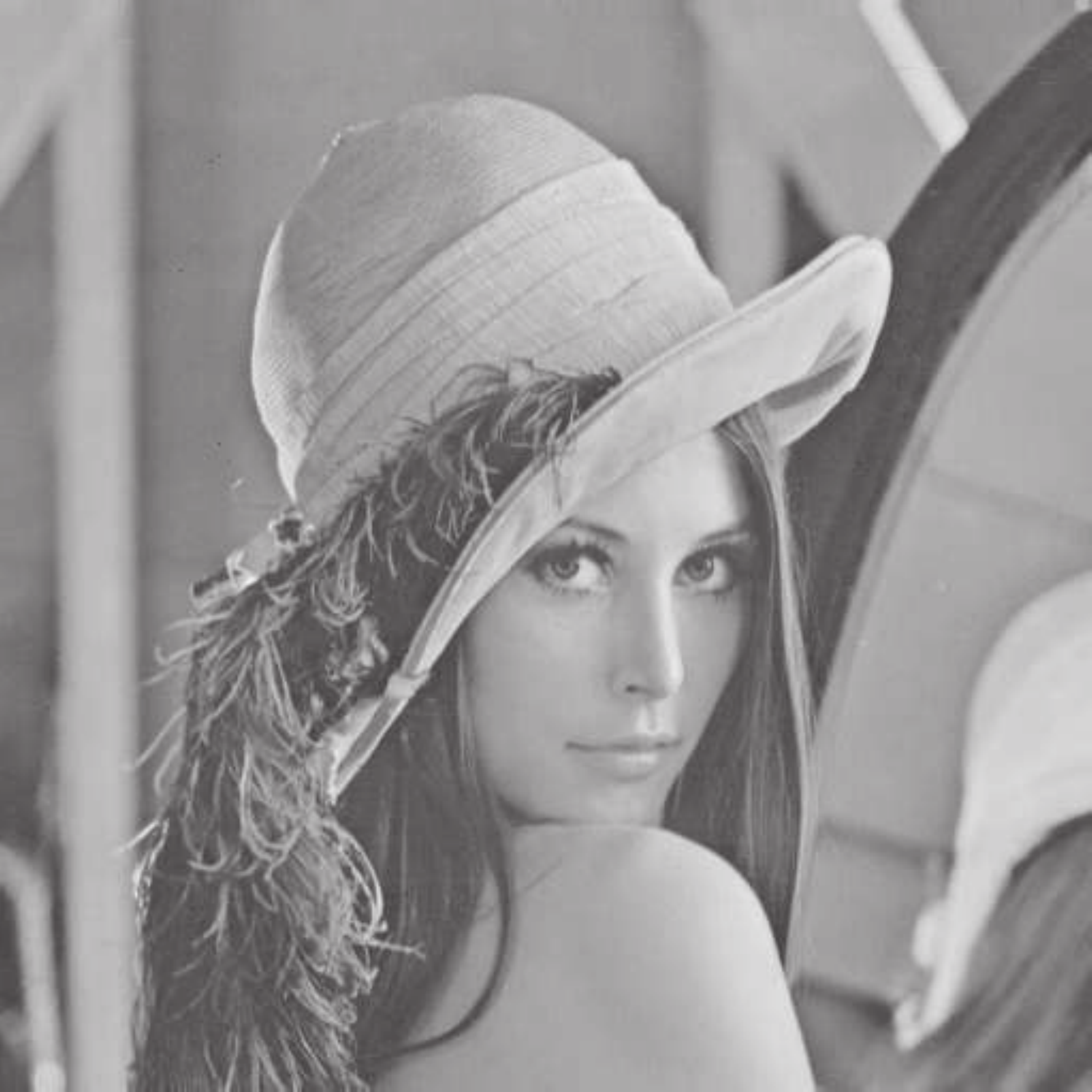}
  }
   \subfigure[]{
   \label{fig-8-b}
    \includegraphics[width=0.2\textwidth]{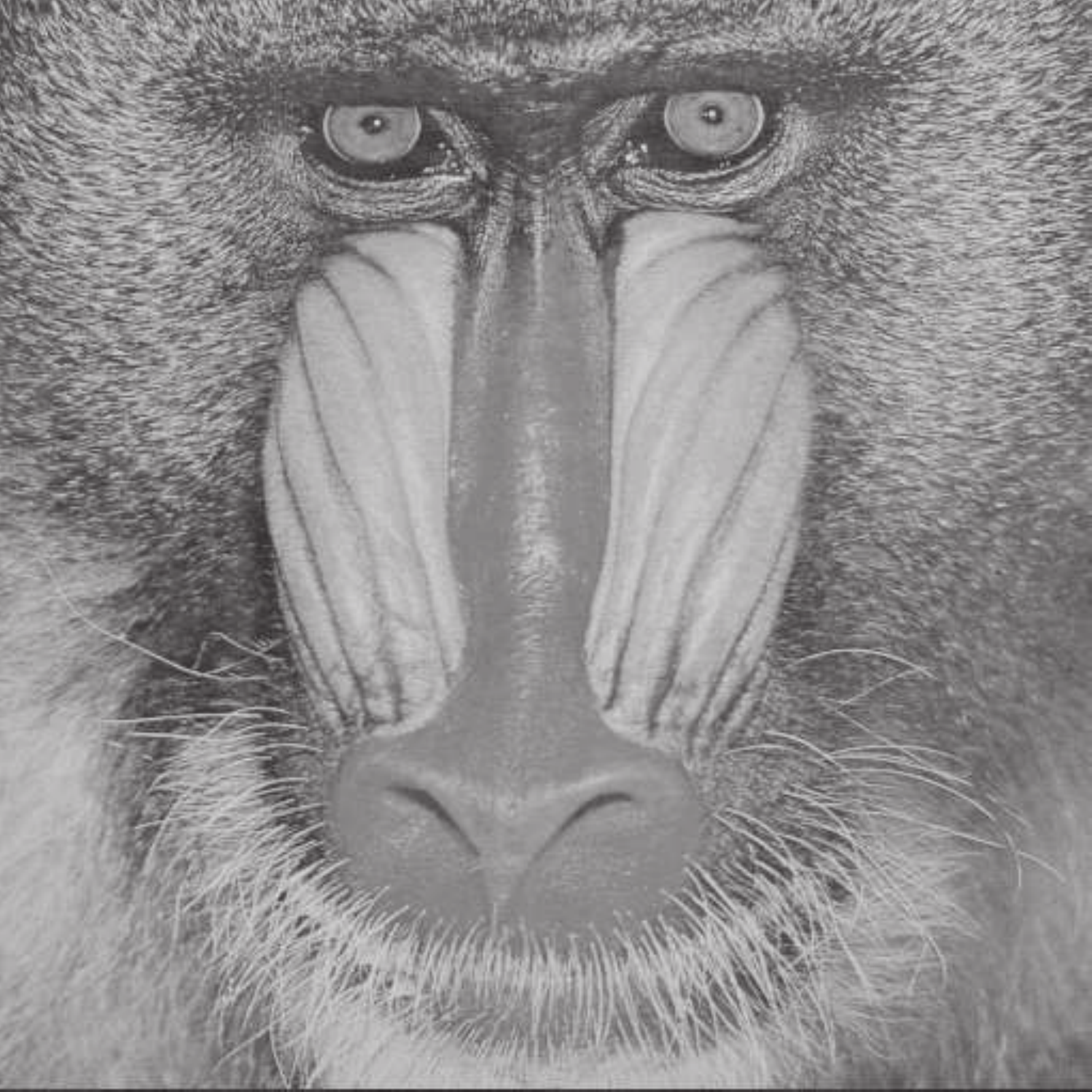}
  }
   \subfigure[]{
   \label{fig-8-c}
    \includegraphics[width=0.2\textwidth]{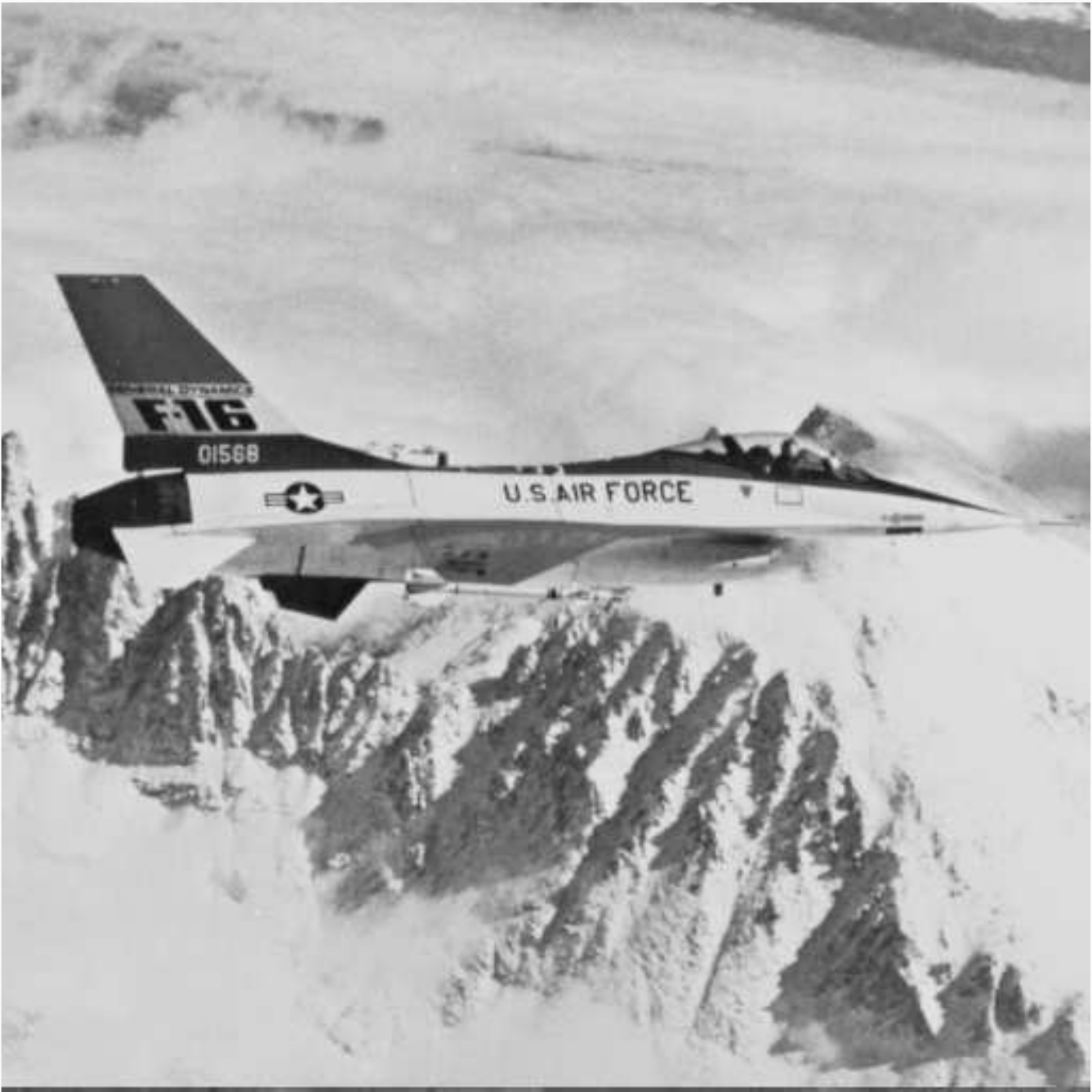}
  }\\
   \subfigure[]{
   \label{fig-8-d}
    \includegraphics[width=0.2\textwidth]{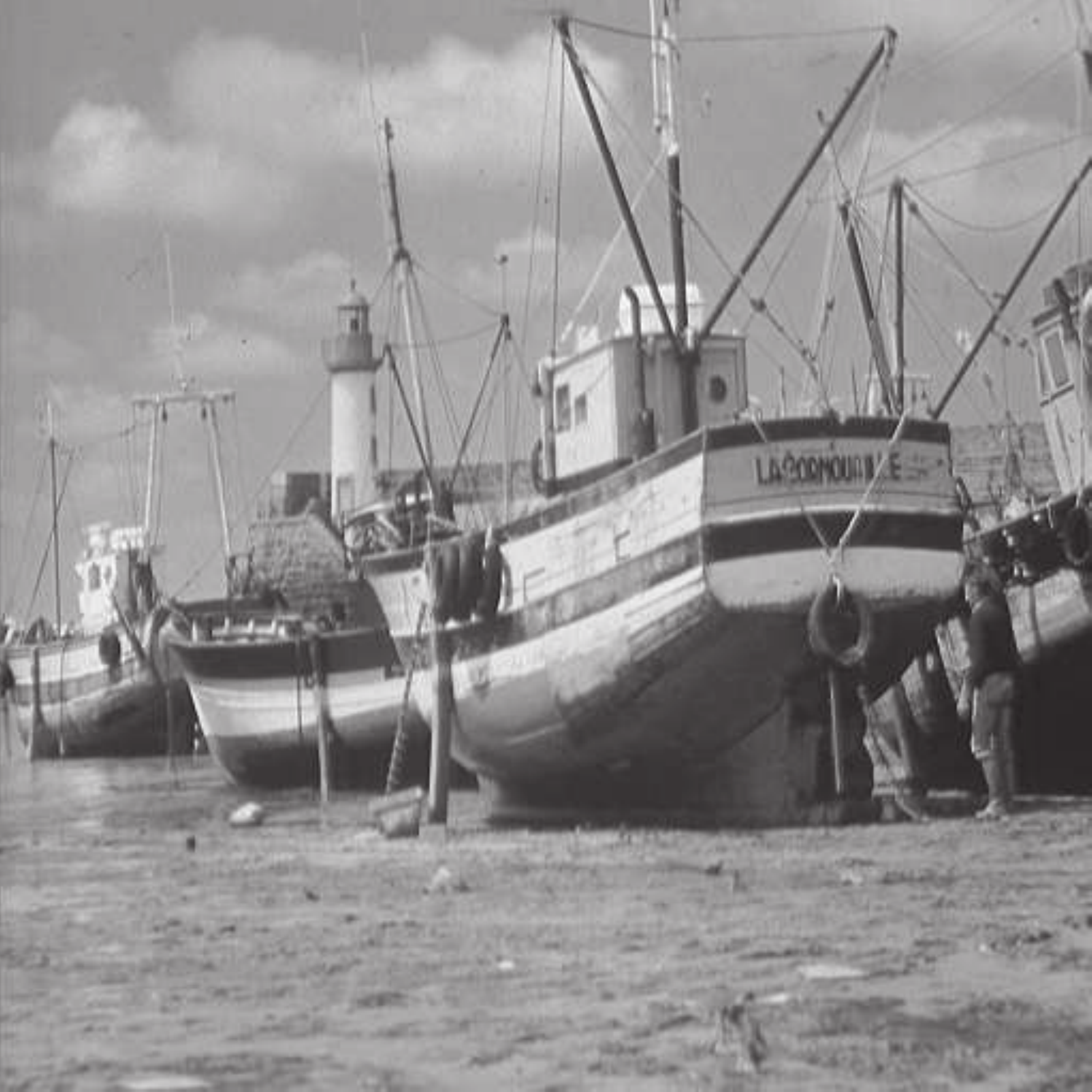}
  }
   \subfigure[]{
   \label{fig-8-e}
    \includegraphics[width=0.2\textwidth]{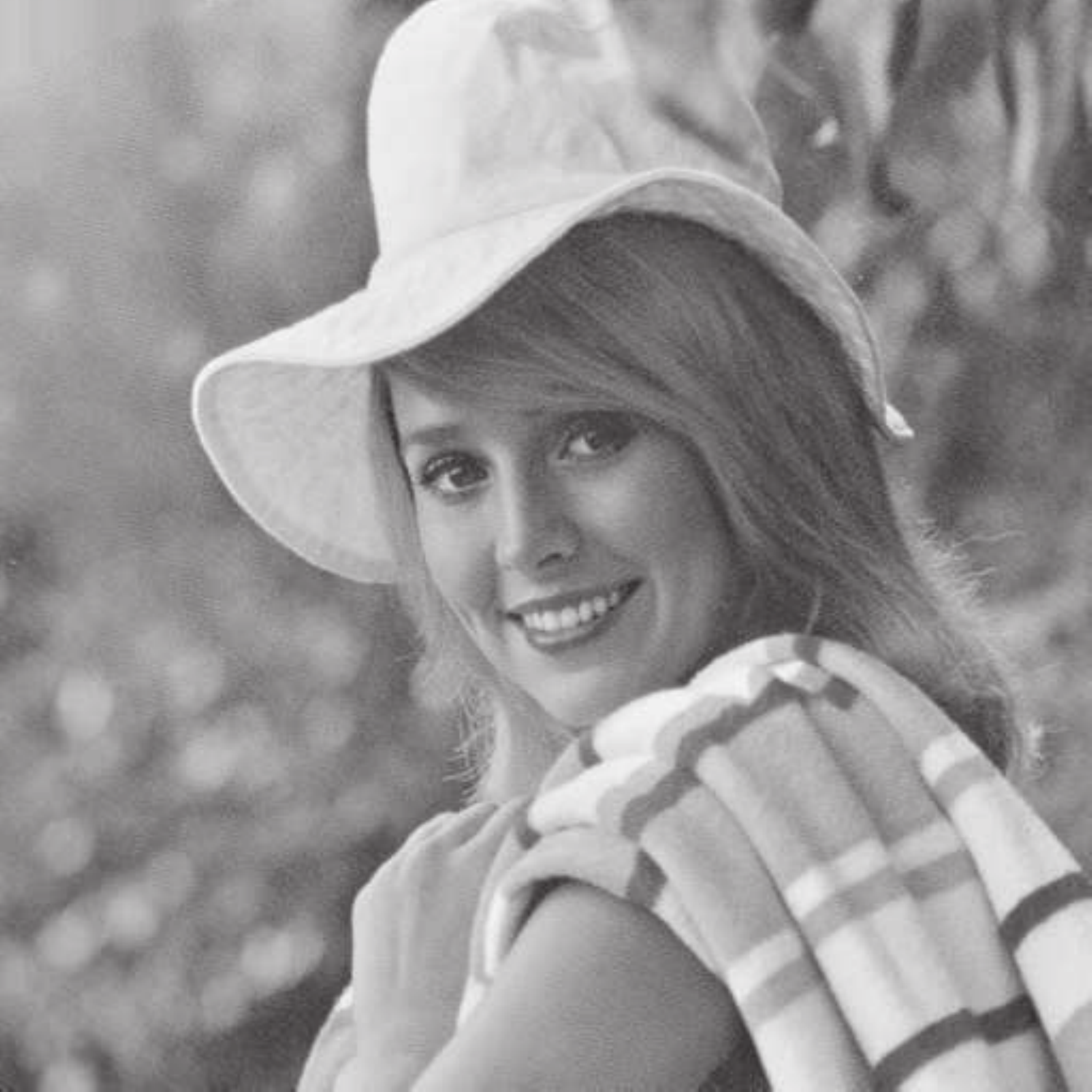}
  }
  \subfigure[]{
  \label{fig-8-f}
    \includegraphics[width=0.2\textwidth]{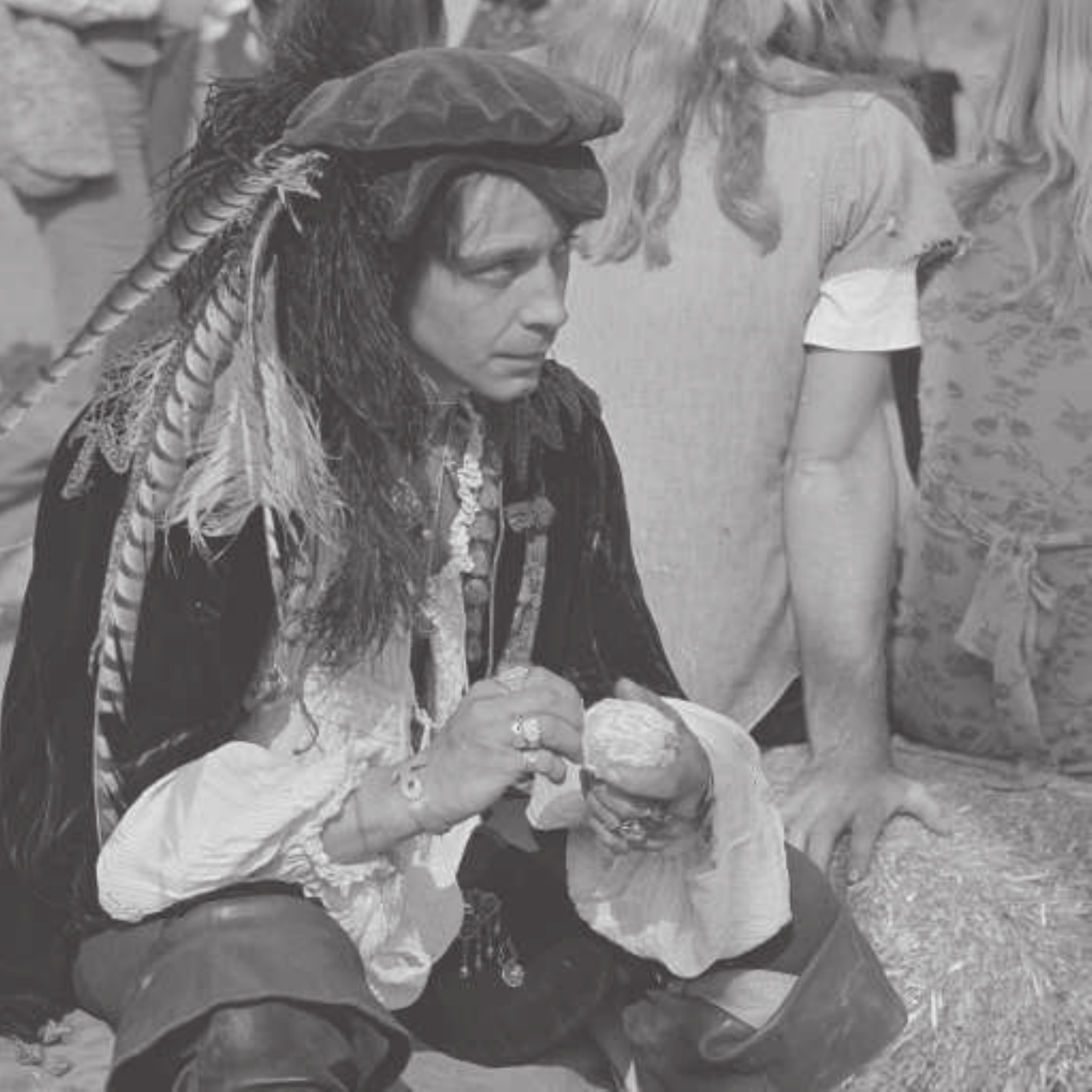}
  }
  \caption{ Test images: (a)Lena. (b)Baboon. (c)Airplane. (d)Boat. (e)Elaine. (f)Man.}
\label{testimage}
\end{figure*}
\begin{figure*}[!ht]
  \centering
    \includegraphics[width=0.75\textwidth]{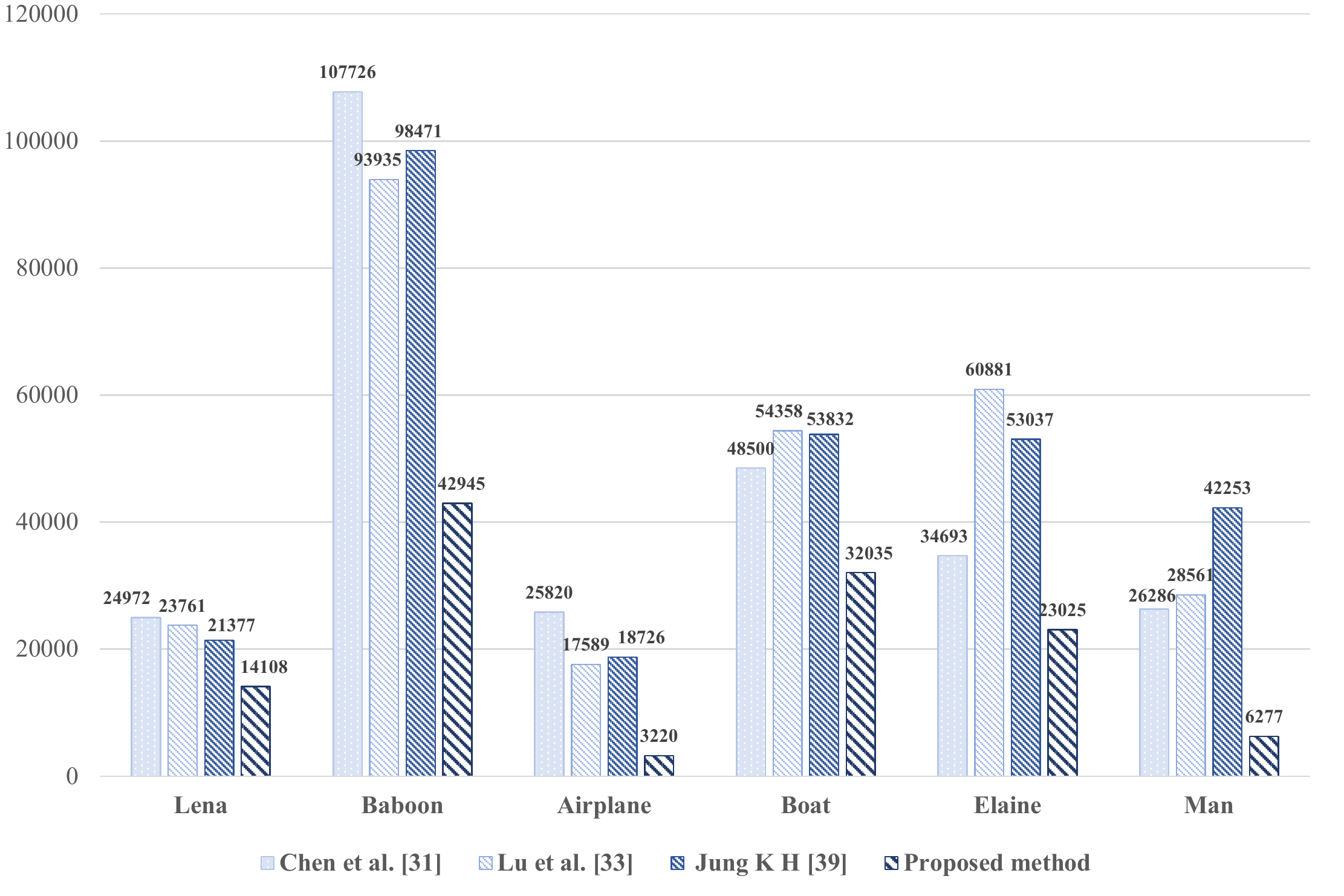}
  \caption{Comparison of the number of ISPs.}
\label{invalid1}
\end{figure*}
\par Extraction and recovery are the reverse process, first extract the additional data from B and recover B, then do the same on A. The detailed process is described below:
\begin{enumerate}
\item Compute the fluctuation values and marked prediction errors of all pixels in B according to ~\ref{subsec::fluctuation values} and ~\ref{subsec::prediction error}. It should be noted that because the fluctuation values of B are calculated using the pixels in A, the fluctuation value here is the same as that before B is embedded. The fluctuation value sequence and the marked prediction error sequence are obtained in the same scanning order as the embedding. The sequence of fluctuation values is arranged in ascending order to obtain $\left \{ F _{B_{1}},F _{B_{2}},...,F _{B_{i}} \right \}$, sorting the prediction error sequence according to the ascending order of the fluctuation value to obtain the marked prediction error sequence $\left \{ {e}'_{F _{B_{1}}},{e}'_{F _{B_{2}}},...,{e}'_{F _{B_{i}}} \right \}$.
\item The additional data in sequence $\left \{ {e}'_{F _{B_{1}}},{e}'_{F _{B_{2}}},...,{e}'_{F _{B_{i}}} \right \}$ is sequentially extracted until half of the payload is reached.The way to extract additional data is as follows:
\begin{small}
\begin{equation}
\centering
\label{eq::eq11}
\ b\! =\!
\left\{
\begin{array}{ll}
0 & {\text{ if } e_{i}^{'} = PK_{1}\ \! or \ e_{i}^{'}\!=\!PK_{2}} \\
\multirow{2}*{$1$} & \text{ if } e_{i}^{'} = \text{min}\left \{PK_{1}\!,\! PK_{2} \right \}\! -\! 1\text{,} \  \\& or \ \! e_{i}^{'}\! =\! \text{max}\left \{PK_{1}\! ,\!PK_{2}\right \} \!+\! 1
\end{array}\right.
\end{equation}
\end{small}
\par where $PK_{1}$, $PK_{2}$, and $Z_{1}$, $Z_{2}$ are the same as that in the procedure of embedding. For the extraction of data, b is extracted when ${e}'_{i}\epsilon\left\{\begin{matrix}
min \left (PK_{1},PK_{2}\right)-1,max\left(PK_{1},PK_{2} \right)
\end{matrix}\right.\\\left.\begin{matrix}
+1,PK_{1},PK_{2}
\end{matrix}\right\}$.
\item In the process of extracting additional data and recovering the original image, the prediction error is recovered as follows:
\begin{small}
\begin{equation}
\label{eq::eq12}
\centering
\ e_{i}\! =\left\{\! \begin{array}{ll}
 e_{i}^{'}-1 & \text{ if } e_{i}^{'}=max\left ( PK_{1},PK_{2} \right )+1 \\
 e_{i}^{'}+1 & \text{ if } e_{i}^{'}=min\left ( PK_{1},PK_{2} \right )-1 \\
 e_{i}^{'} & \text{ if } e_{i}^{'}=PK_{1}\ or\ e_{i}^{'}=PK_{2}  \\
 \multirow{2}*{${e}'_{i}-1$} & \text{ if } e_{i}^{'}\! \leq \! \text{max}\left ( Z_{1},Z_{2} \right )\ \ \ \ \ \ \ \ \ \ \ \ \ \ \ \ \ \ \text{.} \\& and  \ e_{i}^{'}\! >\!  \text{max}\left ( PK_{1},PK_{2} \right ) \\
 \multirow{2}*{${e}'_{i}+1$} & \text{ if } e_{i}^{'}\! \geq \! \text{min}\left ( Z_{1},Z_{2} \right )\ \\& and  \ e_{i}^{'}\! <\!  \text{min}\left ( PK_{1},PK_{2} \right )\\
 e_{i}^{'}& \text{ otherwise }
\end{array}\right.
\end{equation}
\end{small}
\par Correspondingly, the original pixel values are reconstructed as:
\begin{equation}
\label{eq::eq13}
\centering
\ P_{i}=P_{i}^{''}+e_{i}\text{,}
\end{equation}
\par where $P_{i}^{''}$ is the predicted pixel value, $e_{i}$ is the corresponding recovered prediction error, and $P_{i}$ is the recovered pixel value.
\item After the extraction and recovery of B is done, the extraction and recovery of A is completed in the same way based on B. In addition, the pixel value is recovered based on the decompressed location map. Change 254 to 255 and 1 to 0 when the location map is marked 1. Finally, the entire image is completely recovered.
\end{enumerate}
\par The examples of histogram shifting in embedding and extraction are shown in Fig.~\ref{emd-ext}. It depicts the case of histogram shifting that embeds and extracts additional data when $PK_{2}>PK_{1}$, $Z_{2}> Z_{1}$.

\section{Experimental results and analysis}
\label{sec::Experimental}
\par In this section, we will evaluate the performance of the proposed method through several experiments and analyze the results of the experiment. Six standard $512\times512$ sized test images, Lena, Baboon, Airplane, Boat, Elaine, Man, are used in the experiments, as shown in Fig.~\ref{testimage}. We compare with three state-of-the-art schemes of Chen et al.~\cite{chen2013reversible}, Lu et al.~\cite{lu2016asymmetric}, and Jung K H~\cite{jung2017high} to estimate the performance of the proposed method in terms of image visual quality and embedding capacity.

\subsection{Comparison with other schemes}
\label{subsec::other scheme}
\par The schemes of Chen et al.~\cite{chen2013reversible}, Lu et al.~\cite{lu2016asymmetric}, and Jung K H~\cite{jung2017high} are chosen for comparing with the proposed scheme. For~\cite{chen2013reversible}, Chen et al. used maximum prediction error and minimum prediction error to form an asymmetric prediction error histogram. Thus, the image quality is improved with pixel complementary methods. Lu et al.~\cite{lu2016asymmetric} improved on the basis of Chen et al.’s scheme and enhanced the accuracy of the prediction error combining the prediction methods of Feng et al.’s~\cite{feng2012reversible} and Lukac et al.’s~\cite{lukac2004digital} , so there is a big improvement in image quality and embedding capacity. In~\cite{jung2017high}, Jung K H sorted the pixels in the block after dividing the image into a number of $3\times 1$ non-overlapping blocks, and the maximum and minimum pixels were used to embed the additional data and the image was recovered by PEE. In our scheme, exploiting the fluctuation combined with PEH, the additional data is preferentially embedded in the position where the fluctuation value is small, hence, effectively reducing the number of ISPs and performance is greatly improved.

\subsubsection{Comparison of the number of ISPs }
\label{subsubsec::invalid}
\par Peak signal-to-noise-ratio (PSNR) is often used as an objective measure of image quality, which reflects the degree of similarity between the stego image and the original image. For grayscale images, it is defined as follows:
\begin{equation}
\label{eq::eq14}
\centering
\ \text{PSNR}=10\times \text{log}_{10}\frac{255^{2}}{\text{MSE}}\text{,}
\end{equation}
where MSE refers to the mean square error between the original image and the stego image. For a test image size of $512\times512$, MSE is defined as:
\begin{equation}
\label{eq::eq15}
\centering
\ \text{MSE}=\frac{1}{512\times512}\sum_{i=1}^{512}\sum_{j=1}^{512}(o_{ij}-s_{ij})^{2}\text{,}
\end{equation}
where $o_{ij}$ is the original pixel value and $s_{ij}$ is the marked pixel value. Furthermore, the MSE between the original image and the stego image can be regarded as the sum of the $\text{MSE}_{1}$ generated by the valid shifting pixel and the $\text{MSE}_{2}$ generated by the ISPs, expressed as:
\begin{equation}
\label{eq::eq16}
\centering
\ \text{MSE}=\text{MSE}_{1}+\text{MSE}_{2}\text{.}
\end{equation}
\par Given a string of additional data, the number of valid shifting pixels is fixed, and the corresponding $\text{MSE}_{1}$ is also fixed. Therefore, the more the number of ISPs, the larger the corresponding $\text{MSE}_{2}$, resulting in larger MSE. Correspondingly, the smaller the PSNR, the larger the image distortion. In this paper, we reduce the invalid shifting of the pixels by preferentially embedding additional data into smooth pixels. In addition, in order to illustrate that the proposed method can effectively reduce the invalid shifting of pixels, Fig.~\ref{invalid1} gives the comparison of the number of ISPs between the other three methods and our proposed method on six text images when additional data is 10000 bits. It can be seen intuitively from the Fig.~\ref{invalid1} that the number of ISPs of the proposed scheme is much less than that of the other three methods on the six test images. For the smooth image, taking the Airplane as an example, the ISPs of the proposed method is 3,220, which is  22600, 14369, and 15506 less than the three method, respectively. For the unsmooth image, Baboon is taken as an example. The method has fewer ISPs than half of the three methods.
\begin{figure*}[!ht]
  \centering
  \subfigure[]{
    \includegraphics[width=0.4\textwidth]{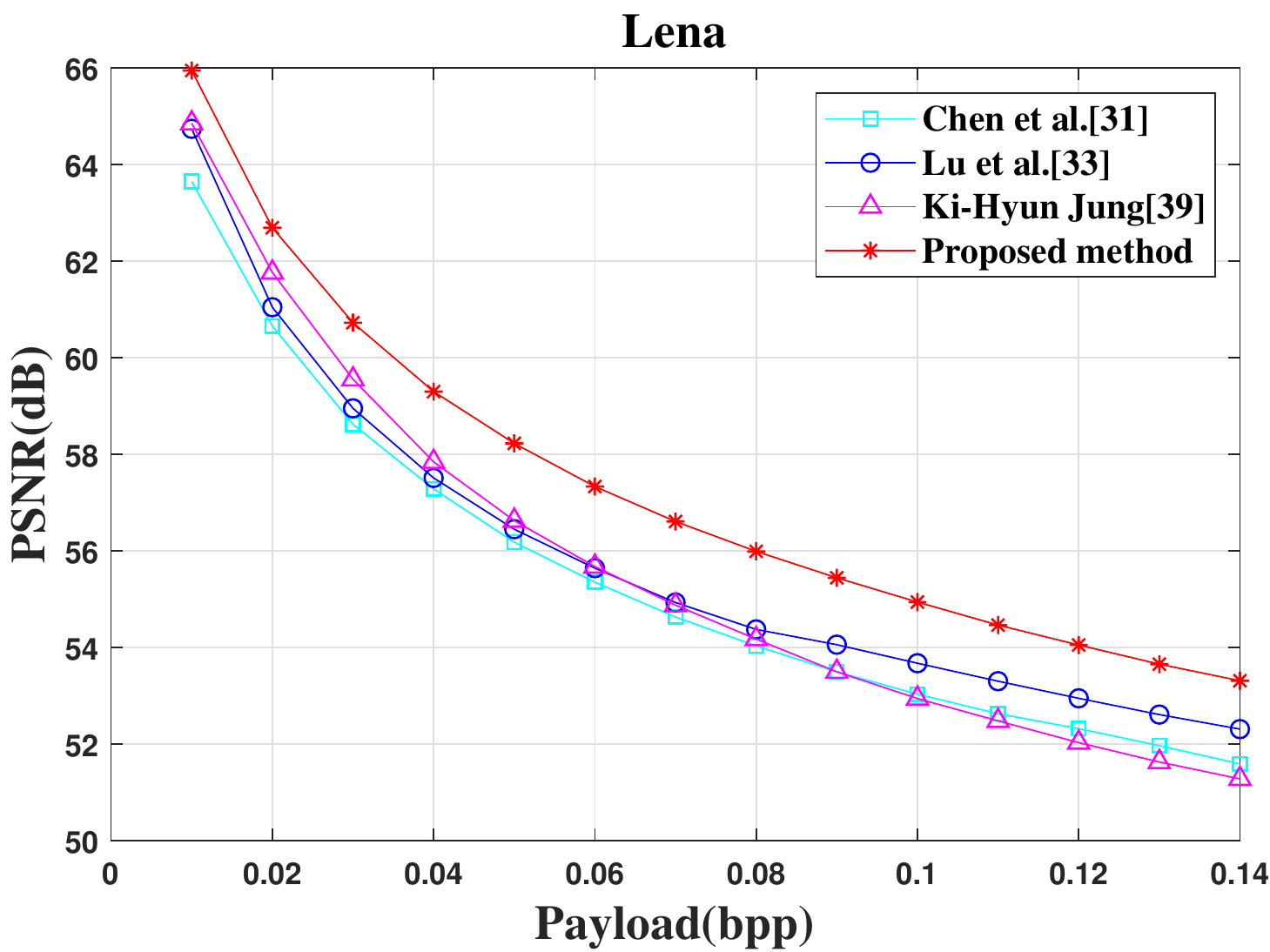}
  }
  \subfigure[]{
    \includegraphics[width=0.4\textwidth]{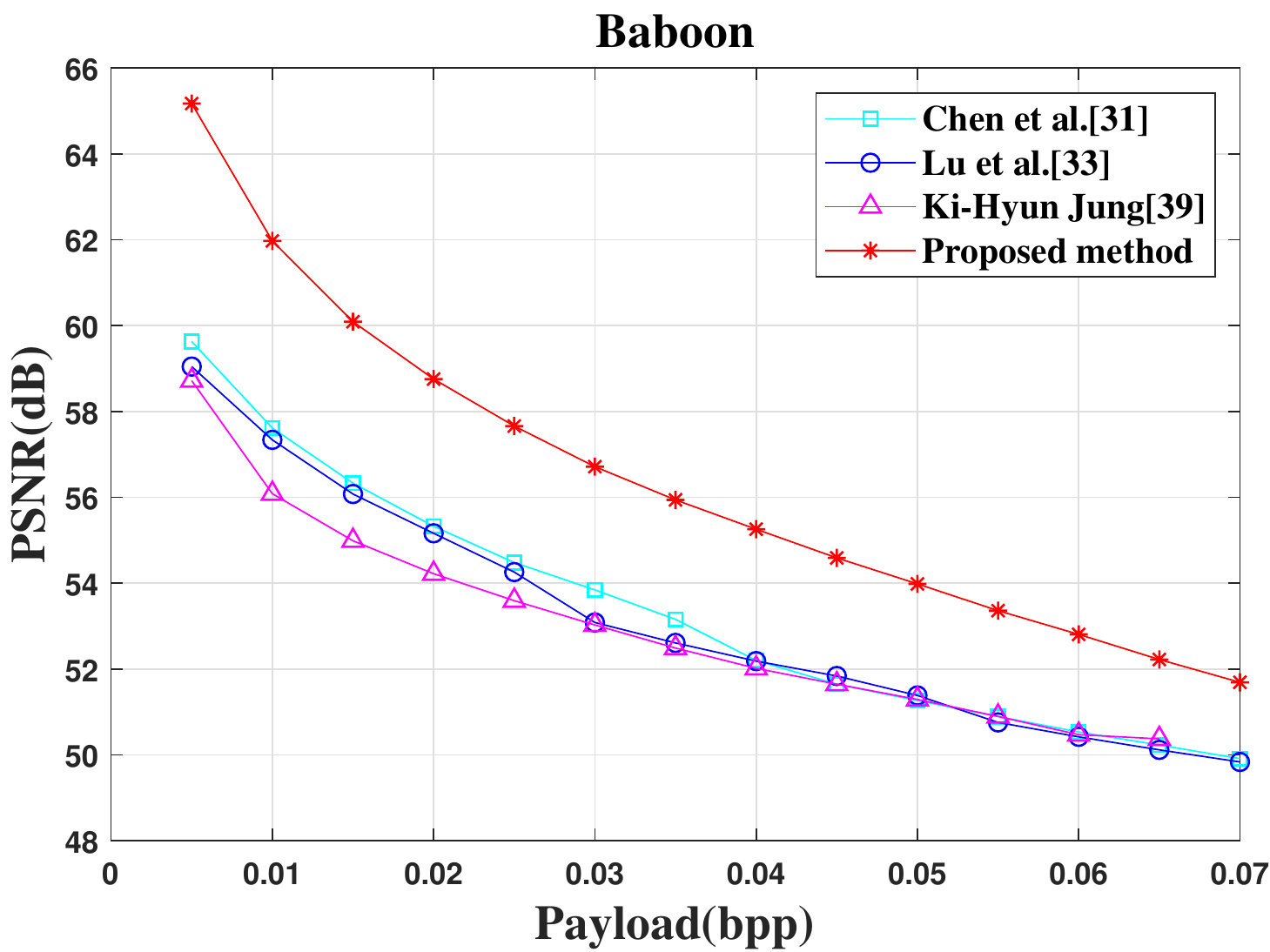}
  }
   \subfigure[]{
    \includegraphics[width=0.4\textwidth]{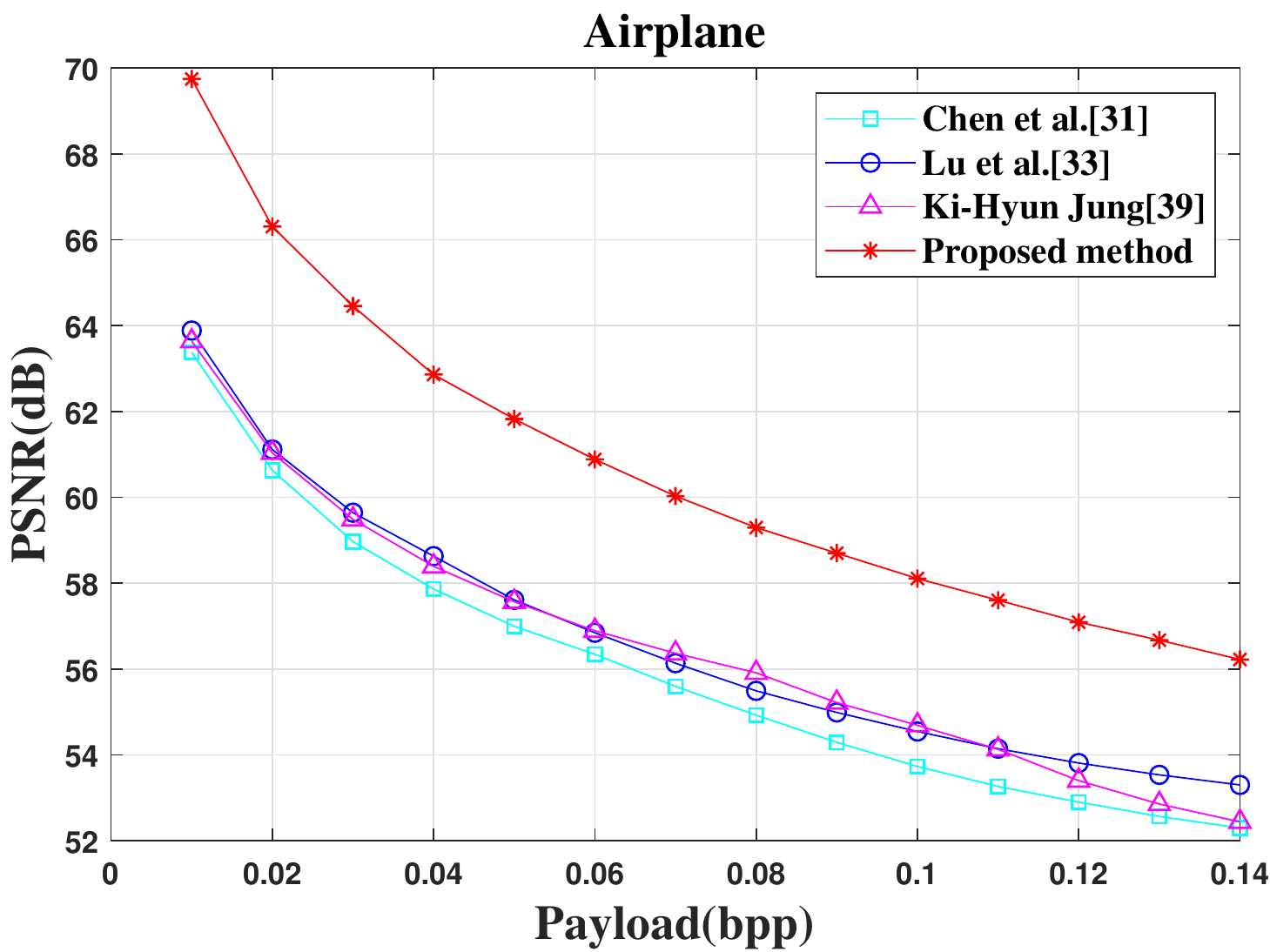}
  }
   \subfigure[]{
    \includegraphics[width=0.4\textwidth]{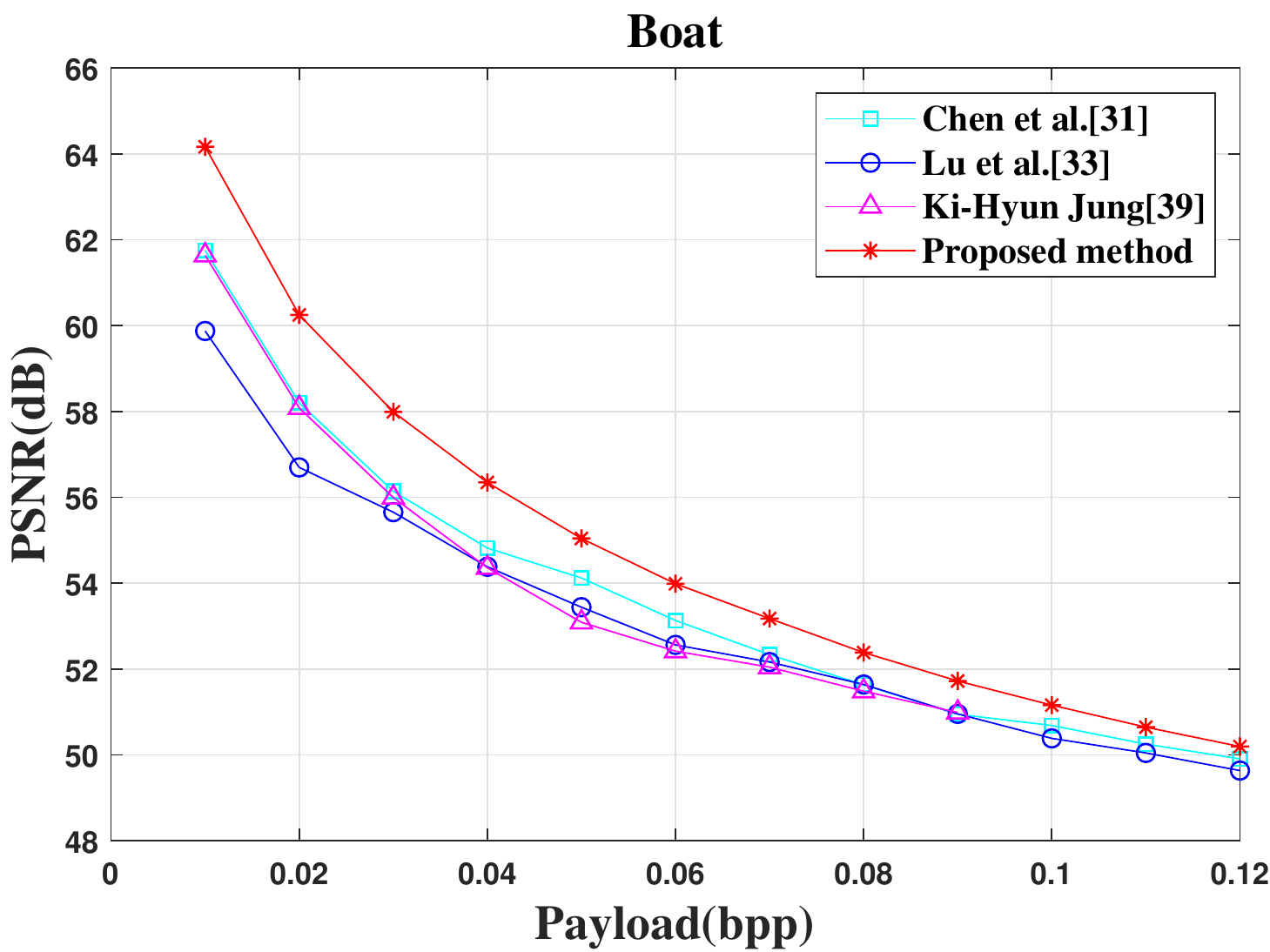}
  }
   \subfigure[]{
    \includegraphics[width=0.4\textwidth]{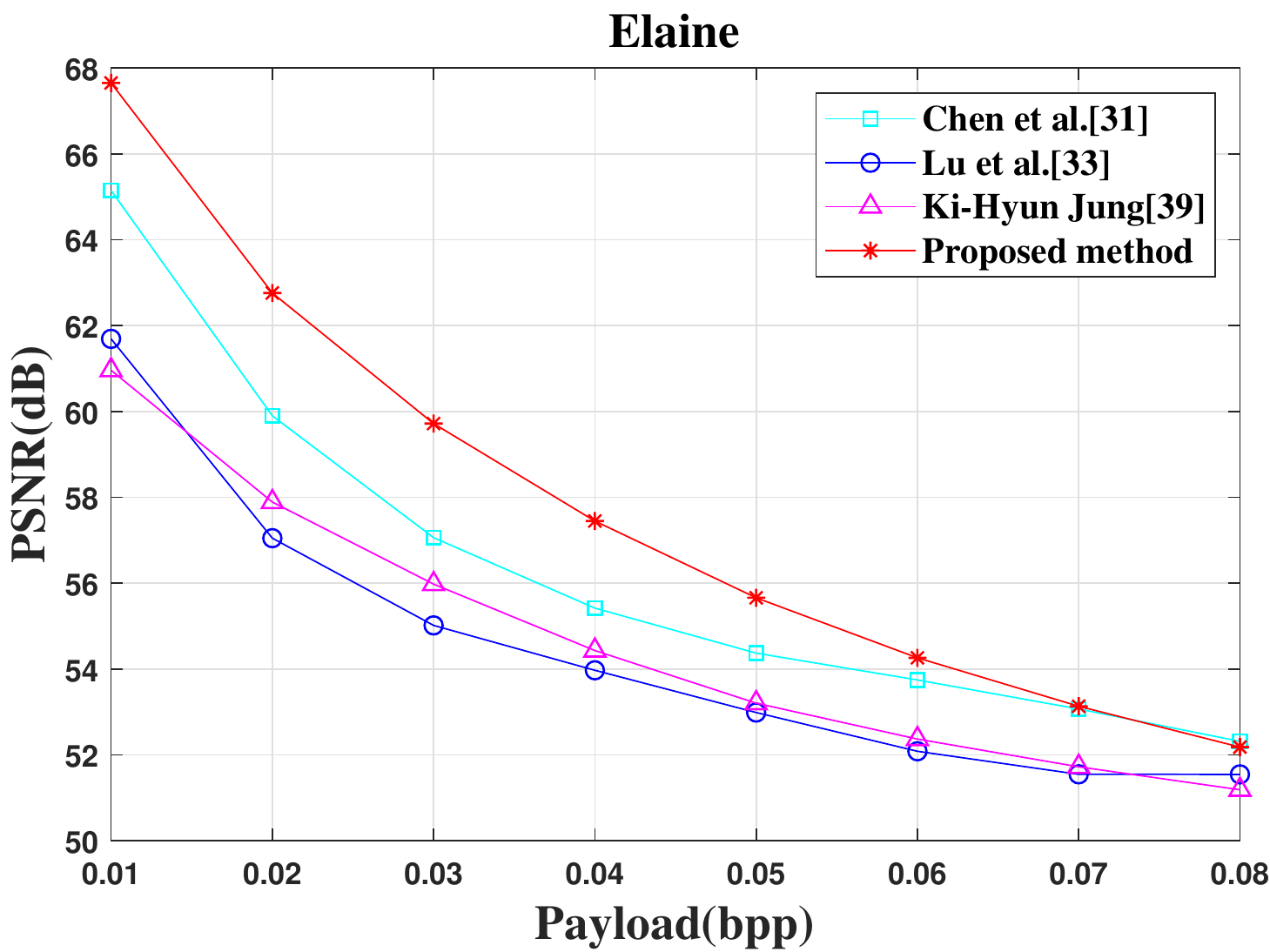}
  }
   \subfigure[]{
    \includegraphics[width=0.4\textwidth]{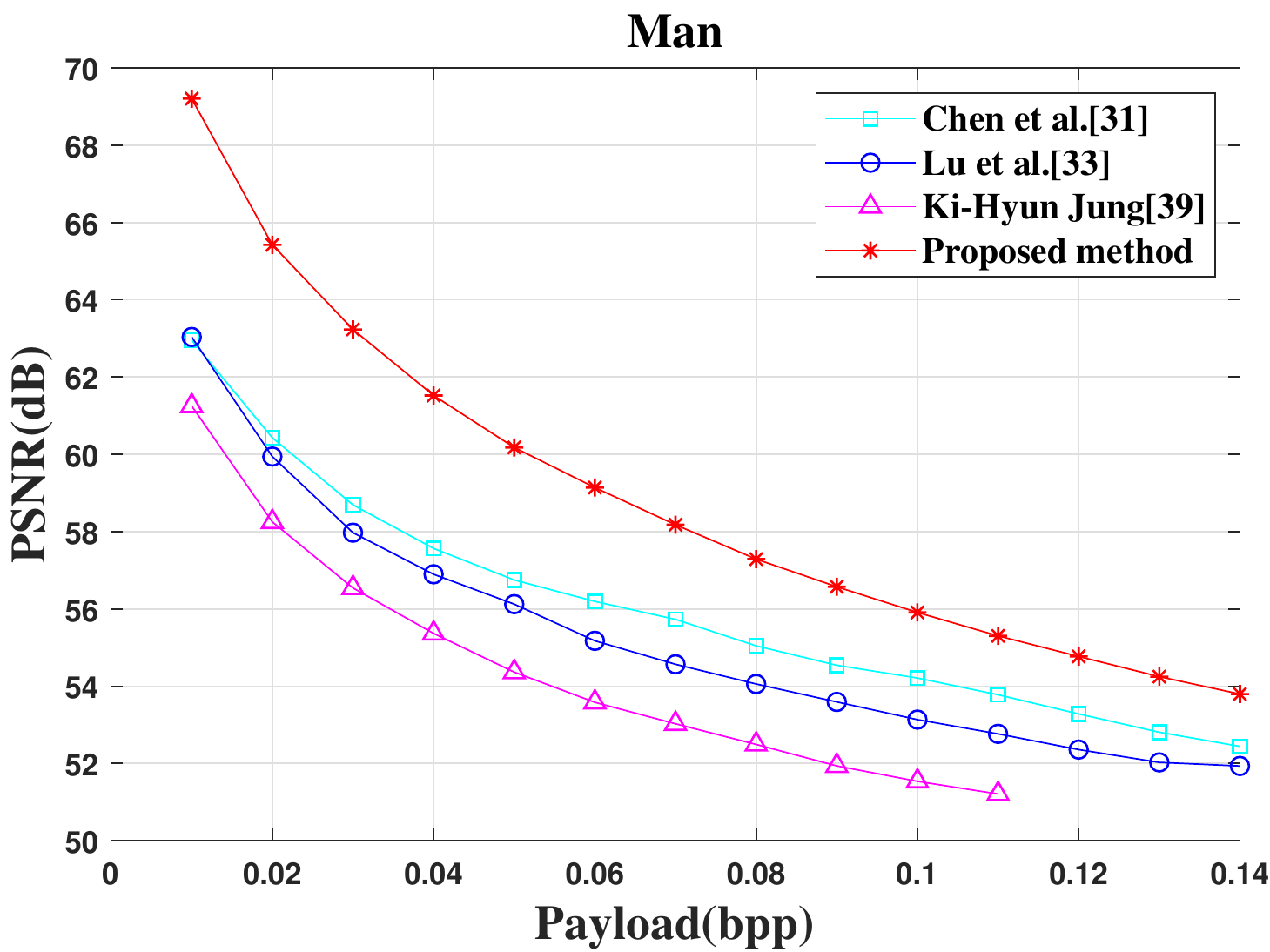}
  }
  \caption{ Comparison of the performance with other schemes.}
  \label{psnr}
\end{figure*}
\begin{table*}[!ht]\scriptsize
\caption{\label{tb::tab1} The average PSNR gains (in dB) for the payload 0.01$\sim$0.05 (bpp) by comparing with three schemes.}
\centering
\begin{tabular}{llll}
\hline
Image    & Gain vs Chen et al.~\cite{chen2013reversible} & Gain vs Lu et al.~\cite{lu2016asymmetric}  & Gain vs Jung K H~\cite{jung2017high} \\ \hline
Lena     & 2.10                         & 1.64                       & 1.25                      \\
Baboon   & 3.29                         & 3.48                       & 4.00                      \\
Airplane & 5.47                         & 4.86                       & 5.01                      \\
Boat     & 1.75                         & 2.74                       & 2.12                      \\
Elaine   & 2.27                         & 4.50                       & 4.16                      \\
Man      & 4.63                         & 5.12                       & 6.76                      \\ \hline
\end{tabular}
\end{table*}
\begin{figure*}[!ht]
  \centering
    \includegraphics[width=0.75\textwidth]{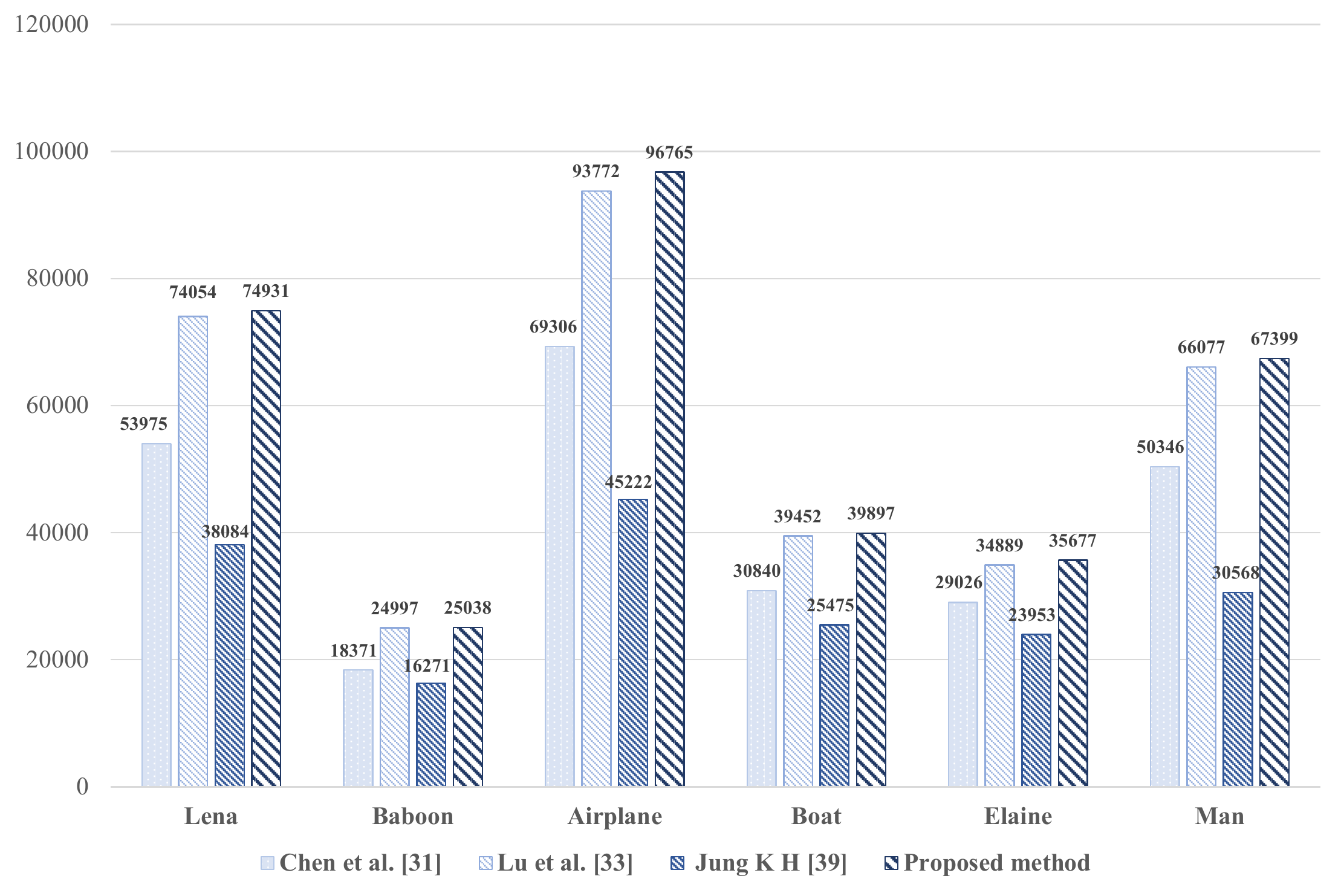}
  \caption{Comparison of capacity for each image by using four schemes.}
\label{Capacity1}
\end{figure*}

\subsubsection{Comparison of PSNR}
\label{subsubsec::PSNR}
\par Fig.~\ref{psnr} is a comparison of the performance of our scheme with other schemes. Here, combined with the image characteristics, we set the payload step size to 0.01(bpp) or 0.02(bpp) to compare the performance. In addition, since all schemes are embedded at one time, for the sake of fairness, Lu et al.'s~\cite{lu2016asymmetric} scheme is embedded only once. It can be seen from Fig.~\ref{psnr} that the proposed scheme has better performance under different payloads than Chen et al.~\cite{chen2013reversible}, Lu et al.~\cite{lu2016asymmetric} and Jung K H~\cite{jung2017high}. Taking the texture image Baboon and the smooth image Airplane as example, the PSNR of the proposed scheme is greater than 58 dB when the embedding capacity of baboon is 0.02 (bpp), but the PSNR of the schemes of Chen et al., Lu et al. and Jung KH cannot reach 56 dB. For Airplane, when the embedding capacity is 0.03 (bpp), the proposed schemes, Chen et al., Lu et al. and Jung K H, have PSNR of 64.45, 58.97, 59.64, and 59.49 dB, respectively. Our scheme outperforms~\cite{chen2013reversible},~\cite{lu2016asymmetric} and~\cite{jung2017high} with an increase of PSNR by 5.48, 4.81 and 4.96 dB.
\par Further, Table~\ref{tb::tab1} demonstrates the average gains of PSNR when the embedding capacity is 0.01$\sim$0.05 (bpp) compared with the above three schemes. As seen in the table that compared with the schemes of Chen et al.~\cite{chen2013reversible}, Lu et al.~\cite{lu2016asymmetric} and Jung K H~\cite{jung2017high}, the proposed scheme provides a consistent gain in different test images. In particular, compared to~\cite{jung2017high}, the gain of the test image Man reaches 6.76. Combined with the given Fig.~\ref{psnr} and Table~\ref{tb::tab1}, we can find that the proposed method can reduce the distortion to some extent compared with the other three methods, whether it is smooth image or unsmooth image.

\subsubsection{Comparison of embedding capacity}
\label{subsubsec::capacity}
\par The comparison of embedding capacity between the proposed scheme and Chen et al.'s~\cite{chen2013reversible}, Lu et al.'s~\cite{lu2016asymmetric}, Jung K H's~\cite{jung2017high} schemes in six test images is displayed in Fig.~\ref{Capacity1}. It can be obtained intuitively from the table that the proposed scheme effectively improves the embedding capacity. For the texture image Baboon, the embedding capacity is 18371 bits, 24997 bits, 16271 bits respectively in~\cite{chen2013reversible},~\cite{lu2016asymmetric},~\cite{jung2017high}, while the proposed scheme is 25038 bits. Compared with the three schemes, the proposed scheme increased by 36.29\%, 0.16\% and 53.88\%, differently. For the smooth image Lena, the proposed scheme is 38.83\%, 0.33\%, and 96.75\% higher than the other three schemes. This is because the proposed scheme selects double peak, which effectively increases the embeddable position in the image. Here, it should be noted that the embedding capacity of Chen et al.~\cite{chen2013reversible}, Lu et al.~\cite{lu2016asymmetric} and the proposed method may vary slightly due to the embedded additional data, but the variation is very small.This is because these three methods rely on the value of the stego pixel during the embedding process. In addition, it can be seen that the method in~\cite{lu2016asymmetric} has a small gap between the embedded capacity and the proposed method because the image is divided into two parts in~\cite{lu2016asymmetric}, and each part is embedded twice.

\section{Conclusions}
\label{sec::Conclusion}
\par In this paper, we propose a RDH scheme to reduce the number of ISPs combined with image texture. The image is divided into two parts by the checkerboard pattern, the fluctuation value of each pixel in each part is calculated to determine the degree of smoothness, and the fluctuation value is sorted in ascending order. Furthermore, combined with the prediction error, the additional data is preferentially embedded in the position of the smooth pixel, thereby effectively reducing the invalid shifting of pixels. It is the advantage of using checkerboard method that makes the auxiliary information less. The improvement of experimental performance is divided into two parts. First of all, compared with three state-of-the-art schemes~\cite{chen2013reversible},~\cite{lu2016asymmetric},~\cite{jung2017high}, due to the preference for smooth pixel point while embedding, our PSNR has been significantly improved, and good image visual quality is acquired. Secondly, our embedding capacity has been greatly raised by selecting double peak. In future research, we will further study the performance differences between smooth images and texture images, explore the characteristics of smooth image and texture image, and design more effective methods to reduce the number of ISPs for smooth image and non-smooth image respectively.
%\section*{References}
%\section*{References}
%\bibliographystyle{splncs03}
%\bibliography{reference}

\bibliography{mybibfile}

\par{
\vspace{4em}
\begin{wrapfigure}{l}{25mm}
    \includegraphics[width=1in,height=1.25in,clip,keepaspectratio]{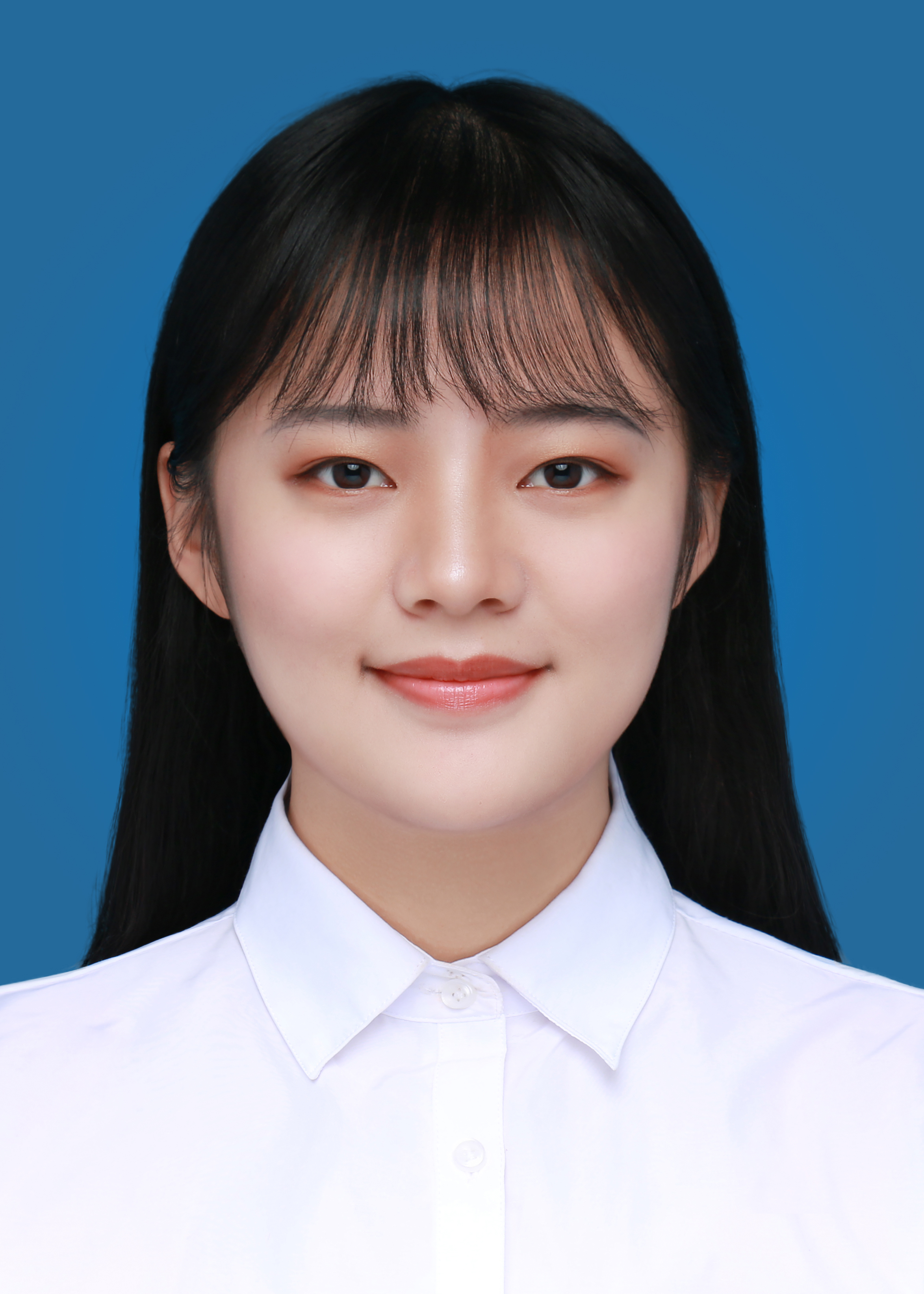}
  \end{wrapfigure}
  \par \textbf{Yujie Jia} received her bachelor degree in electronics and information engineering in 2016 and now is a master student in the School of Computer Science and Technology, Anhui University. Her current research interests include reversible data hiding.\par}

\vspace{2em}
\par{
\begin{wrapfigure}{l}{25mm}
    \includegraphics[width=1in,height=1.25in,clip,keepaspectratio]{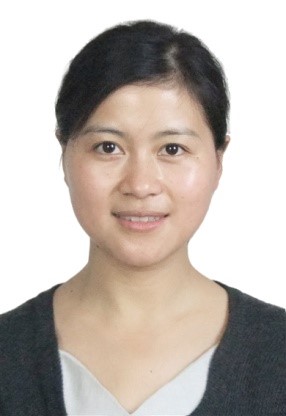}
\end{wrapfigure}\par
\textbf{Zhaoxia Yin} received her B.Sc., M.E. \& Ph.D. from Anhui University in 2005, 2010 and 2014 respectively. She is an IEEE/ACM/CCF member, CSIG senior member and the Associate Chair of the academic committee of CCF YOCSEF Hefei 2016--2017. She is currently working as an Associate Professor and a Doctoral Tutor in School of Computer Science and Technology at Anhui University. She is also the Principal Investigator of two NSFC Projects. Her primary research focus including Information Hiding, Multimedia Security and she has published many SCI/EI indexed papers in journals, edited books and refereed conferences.\par}
\vspace{2em}
\par{
\begin{wrapfigure}{l}{25mm}
    \includegraphics[width=1in,height=1.25in,clip,keepaspectratio]{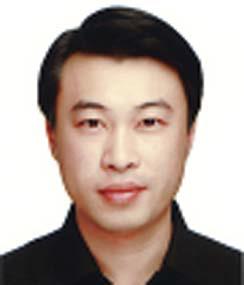}
\end{wrapfigure}\par
\textbf{Xinpeng Zhang} received the B.S. degree in computational mathematics from Jilin University, China, in 1995, and the M.E. and Ph.D. degrees in communication and information system from Shanghai University, China, in 2001 and 2004, respectively. Since 2004, he had been with the faculty of the School of Communication and Information Engineering, Shanghai University, and he is currently a Professor of School of Computer Science, Fudan University. He was with the State University of New York at Binghamton as a visiting scholar from January 2010 to January 2011, and Konstanz University as an experienced researcher sponsored by the Alexander von Humboldt Foundation from March2011 to May 2012. He served IEEE Transactions on Information Forensics and Security as an Associate Editor from 2014 to 2017. His research interests include multimedia security, image processing, and digital forensics. He has published more than 200 papers in these areas.
\vspace{2em}
\par{\begin{wrapfigure}{l}{25mm}
  \includegraphics[width=1in,height=1.25in,clip,keepaspectratio]{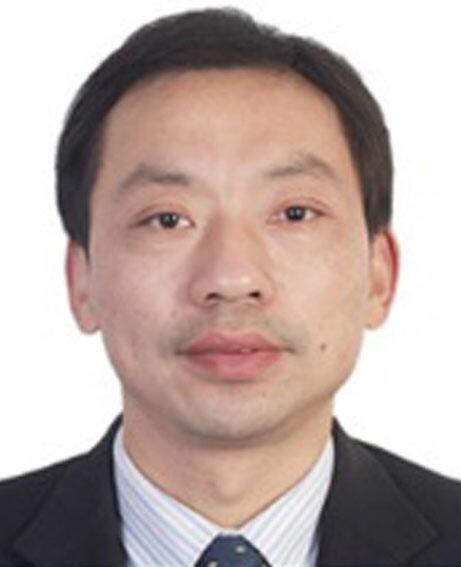}
\end{wrapfigure}\par
\textbf{Yonglong Luo} received his Ph.D. degree from the School of Computer Science and Technology, University of Science and Technology of China in 2005. Since 2007, he has been a professor and a Ph.D supervisor at Anhui Normal University (AHNU). Now he is the Director of Anhui Provincial Key Laboratory of Network and Information Security and the Dean of the School of Computer and Information, AHNU. His main research interests include information security and spatial data processing.\par}
\end{document}